\documentclass[traditabstract]{aa} % for a referee version ,referee
\usepackage{txfonts}
\usepackage{graphics,graphicx}

\begin{document}

\title{%
Why are some galaxy clusters underluminous? The very low concentration
of the CL2015 mass profile
}
\titlerunning{Very low concentration
of the CL2015 mass profile} 
\author{S. Andreon\inst{1}  \and A. Moretti\inst{1}  \and G. Trinchieri\inst{1} \and C. H.
Ishwara-Chandra\inst{2}}
\authorrunning{Andreon et al.}
\institute{
$^1$INAF--Osservatorio Astronomico di Brera, via Brera 28, 20121, Milano, Italy,
\email{stefano.andreon@brera.inaf.it} \\
$^2$National Centre for Radio Astrophysics TIFR, P. B. No. 3., Ganeshkhind, 411007,
Pune, India\\ 
}
\date{Accepted ... Received ...}
\abstract{
Our knowledge of the variety of galaxy clusters has been increasing in the last few years thanks to our progress
in understanding the severity of selection effects on samples.
To understand the reason for the observed variety, we study CL2015, 
a cluster ($\log M_{500}/M_\odot=14.39$) easily missed in X-ray selected observational samples.
Its core-excised X-ray luminosity is low for its mass $M_{500}$, 
well below the mean relation for an X-ray selected sample,
but only $\sim1.5\sigma$ below that derived for an X-ray unbiased sample.  
We derived thermodynamic profiles and hydrostatic
masses with the acquired deep Swift X-ray data, and we used archival Einstein, 
Planck, and Sloan Digital Sky Survey data to derive additional
measurements, such as integrated Compton parameter, total mass, and stellar mass. 
The pressure and the electron density profiles of CL2015 are systematically outside
the $\pm 2\sigma$ range of the universal profiles; in particular
the electron density profile is
even lower than the one derived from Planck-selected clusters. 
CL2015 also turns out  to be fairly different
in the X-ray luminosity versus integrated pressure scaling compared to an X-ray
selected sample, but it is a normal object in terms of
stellar mass fraction. 
CL2015's hydrostatic mass profile, by itself or when is considered together
with dynamical masses, shows that
the cluster has an unusual low concentration and an unusual sparsity
compared to clusters in X-ray selected samples.  
The different behavior of 
CL2015 
is caused by its
low concentration. 
When concentration differences are accounted for,  
the properties of CL2015 become consistent with comparison samples.
CL2015 is perhaps the first known cluster with a remarkably low mass concentration 
for which high quality X-ray data exist.  
Objects similar to CL2015 fail to 
enter observational X-ray selected samples 
because of their low X-ray luminosity relative to their mass. 
The different radial dependence of various observables
is a promising way to collect other examples of low concentration clusters.
}
\keywords{  
galaxies: clusters: general --- Galaxies: clusters: intracluster medium --- dark matter 
--- X-rays: galaxies: clusters --- Radio continuum: galaxies --- 
Galaxies: clusters: individual: CL2015, Abell 117, PSZ2G126.72-72.82
}

\maketitle

\section{Introduction}

Galaxy clusters appear to obey tight scaling relations (e.g., Vikhlinin et al. 2006; Pratt et al. 2009) 
and to possess
universal thermodynamic radial profiles (e.g., Arnaud et al. 2010, Sun et al. 2010, 
Ghirardini et al. 2019). 
Thermodynamic radial profiles (electron density, temperature, pressure, and entropy) offer
a unique asset to the study of the properties of the intracluster medium, the dynamical status
of the cluster, cluster formation and evolution, gas cooling, cluster energetics, 
and structure growth, all of which have prompted a large number of articles on these subjects.

It is now recognized that the X-ray selection deeply affects both scaling relations and 
quantities hinged on them,
for example the cosmological parameter estimates: 
at a given mass, brighter-than-average clusters
are easier to select and make part of a sample, 
while fainter-than-average clusters are easily missed (Stanek et al. 2006).  
Pacaud et al. (2007) show that the evolution of the $L_X-T$ scaling of an X-ray selected
sample would be biased if selection effects were neglected, while
Vikhlinin et al. (2009) accounted for the X-ray selection in their cosmological
analysis, but making assumptions on the unseen population. 
Assumptions on the unseen (or poorly represented) cluster population are also
adopted in scaling relation analyses, for example those of Pacaud et al. (2007), Maughan et al. (2012),
Andreon \& Hurn (2013), and Mantz et al. (2016). Assumptions on the unseen population are also needed for samples that are complete in some observables, for example, that of Giles et al. (2017).

The risk associated with making assumptions on an unseen population has
triggered a number of observational programs targeting, or also including, 
clusters more easily missed.
Andreon \& Moretti (2011) exploited the low and stable X-ray background
of X-ray Telescope (XRT) on Swift for follow-ups of a sample of clusters free from the X-ray bias, finding a larger scatter
in X-ray luminosity at a given richness than in X-ray selected samples (accounting for Malmquist
and selection effect corrections for the latter). Several of these clusters have low surface
brightness, which impair their detection in X-ray surveys. A similar effort was repeated 
by Ge et al. (2019) and by Pearson
et al. (2017), the latter using Chandra data on groups and optical luminosity in place of
richness, finding an increased scatter.  Giles et al. (2015) followed up in X-ray a small sample 
of weak-lensing selected clusters.
Andreon et al. (2009) and
Andreon, Trinchieri \& Pizzolato (2011) observed the two most distant clusters 
free from the X-ray selection bias to constrain
the evolution of the $L_X-T$ scaling without making a hypothesis on the unseen population.
Because of the heavy censoring of X-ray selected samples,
constraints derived from 100 X-ray selected clusters
(Giles et al. 2016) are comparable with the one derived for just the 
two high redshift clusters above.  
In parallel, 
clusters selected by their Sunayev-Zeldovich (SZ) signal turned out to also show a larger scatter
than in X-ray selected samples (Planck Collaboration 2011a, 2012a), with some unexpected
outlier clusters with a low X-ray luminosity for their SZ signal (Planck Collaboration 2016). 
These and other efforts led to the discovery of a growing variety of cluster 
properties at a given mass: 
X-ray luminosity and gas fraction have a larger scatter than previously thought (e.g., 
Andreon \& Moretti 2011; Planck collaboration 2011, 2012; Andreon et al. 2016, 2017; 
Giles et al. 2017, Rossetti et al. 2017), and
clusters with low electron density profiles (Andreon et al. 2016),
or of low surface brightness (Andreon et al. 2016, Xu et al. 2018) have been discovered. 

\begin{figure}
\centerline{\includegraphics[width=9truecm]{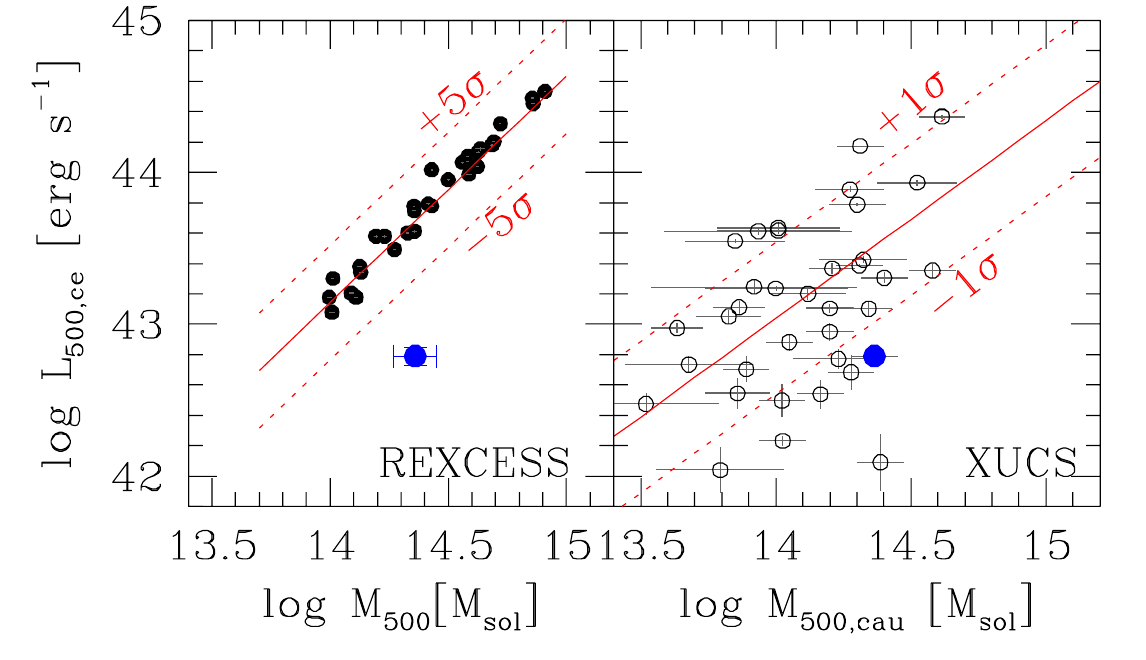}}
\caption{Core-excised [0.5-2] keV band X-ray luminosity vs. mass $M_{500}$.
{\it Left panel:} Black points are REXCESS (X-ray selected) clusters and use $Y_X$-based masses.
The red line is the Malmquist- and selection-bias-corrected fit to the REXCESS X-ray selected sample, 
whereas the dashed lines mark the mean relation $\pm 5\sigma_{intr}$. The blue outlying
point is CL2015. 
{\it Right panel:} Open circles are XUCS clusters, the blue point is CL2015. 
The red line and the $\pm1\sigma$ corridor is a fit to this sample (Andreon et al. 2017). 
}
\label{selLx}
\end{figure}

It is plausible that
X-ray selection effects have consequences on the measured mean thermodynamic 
radial profiles and their scatter. If brighter-than-average clusters are over-represented 
in X-ray samples, then
the average electron density profile is expected to be biased toward the high end because $L_X\propto n^2_e$.
The scatter is expected to be biased toward the low end because a reduced part of the whole population is in the
sample (fainter-than-average clusters are under-represented 
when not missing
altogether). The effect of X-ray selection on the other thermodynamic profiles is
harder to predict, and should be taken from an observational perspective. However, 
obtaining high quality X-ray data for objects missing because they are
intrinsically X-ray faint or of low surface brightness (for their mass) is
observationally hard. This adds to the difficulty
of having accurate masses for a sizeable sample, which is needed to identify outliers at a given mass.

This situation has been partially rectified by the first X-ray unbiased survey of clusters 
(XUCS, Andreon et al. 2016) because these clusters have high quality mass estimates. 
The sample 
consists of a velocity-dispersion-selected sample of clusters in the very
nearby Universe and it is X-ray unbiased at a given mass, because the probability of inclusion
of the cluster in the sample does not depend on its X-ray luminosity (or count rate)
at a fixed mass.
This property allowed a robust estimate of
the scatter (see also Andreon et al 2017a), with a wider range in X-ray luminosity 
(both total and core excised) at a given mass than seen 
in the Representative X-ray selected sample (REXCESS, Pratt et al.
2009, after Malmquist- and selection-bias corrections) and in
the Planck-selected clusters (Andreon et al. 2016). Similar conclusions are also proposed by 
Giles et al. (2017), making assumptions on the extent and size of the unseen population.

The XUCS sample also shows a larger scatter in gas fraction (Andreon et al. 2017). 
While accurate masses and well-determined  X-ray luminosities of XUCS clusters are 
available,  detailed investigation of the X-ray properties of these clusters that are gas poor or 
X-ray faint for their mass requires deeper observations than available. 
In this work, we present the deep X-ray follow-up of CL2015, selected because
it has a low X-ray luminosity for its mass.

Throughout this paper, we assume $\Omega_M=0.3$, $\Omega_\Lambda=0.7$, 
and $H_0=70$ km s$^{-1}$ Mpc$^{-1}$. 
Results of stochastic computations are given
in the form $x\pm y$, where $x$ and $y$ are 
the posterior mean and standard deviation. The latter also
corresponds to 68 \% intervals because we only summarize
posteriors close to Gaussian in this way. All logarithms are in base 10.

\section{Data analyis}

\subsection{Selection of CL2015}

CL2015, also known as Abell 117 and PSZ2G126.72-72.82, is 
an intermediate mass cluster ($\log M_{500}/M_\odot=14.39\pm0.09$) 
in the nearby Universe ($z=0.055$) with a core-excised 
X-ray luminosity of $\log L_{X,500,ce}=42.79\pm0.06$
erg s$^{-1}$
([0.5,2] keV band). This luminosity is entirely consistent with that reported 
by Jones \& Forman (1999) based on the Einstein Imaging Proportional Counter data,  
within an angular aperture close to our $r_{500,cau}$, revised for cosmology, temperature 
($T=3$ keV, see Sect.~3.1 and 3.2), and energy band used here ($\log L_{X}=42.99\pm0.03$ 
vs $42.97\pm0.05$ erg s$^{-1}$ [0.5,2] keV band). The cluster has a richness of $0$ in
the Abell (1958) catalog.

The CL2015 core-excised luminosity is eight times fainter than expected based on its $M_{500}$, 
or about $12\sigma$ below the fit derived using the X-ray selected sample REXCESS (Pratt et al. 2009) after 
corrections applied to account for 
Malmquist and selection biases (see left panel of Fig.~\ref{selLx}). In the analysis of XUCS, 
Andreon et al. (2009) remarked that the sample displays a larger scatter in X-ray luminosity 
at a given mass, relative to REXCESS (or other cluster samples analyzed before).
In this sample, CL2015 is not as extreme, since it is located 
just $\sim1.5\sigma$ away from the average
(right panel of Fig.~\ref{selLx}) and has a $\sim1.5\sigma$
lower gas fraction than average (Andreon et al. 2017). 

This result does not depend on how the total mass is derived.  There are three estimates
of the mass within $\Delta=500$, all consistent with each other:
from the dynamical analysis of Abdullah et al. (2018), $\log M_{500,dyn}=14.27$ (no error quoted),
from the caustic 
technic, $\log M_{500,cau}=14.36\pm0.09$, and from velocity dispersions calibrated with simulations,
$\log M_{500,\sigma_v}=14.27\pm0.12$ (Andreon et al. 2016). Details are 
given in Appendix A and we refer the reader to the original articles for more information on the 
technics used.

CL2015 is one of the many clusters $\sim1.5\sigma$ below the mean XUCS relation and
has been chosen  to minimize exposure time, leading us to select
one of the nearest clusters in XUCS and one of the only two clusters, out of a sample
of 34, with $r_{500}$ slightly larger than the Swift field of view.

\begin{figure}
\centerline{\includegraphics[width=9truecm]{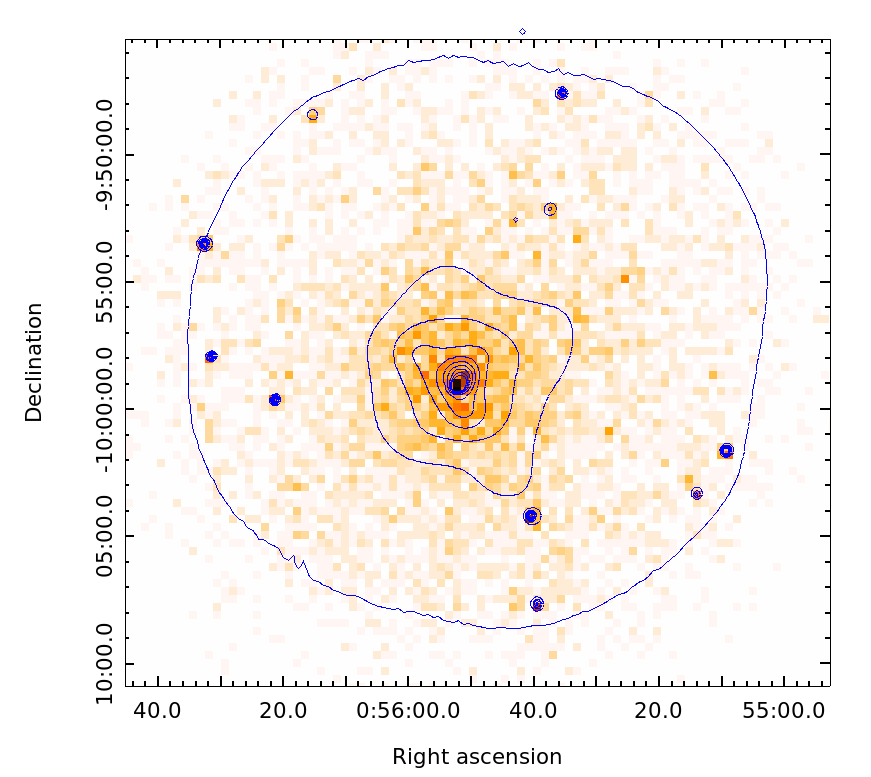}}
\caption{X-ray contours from an adaptively smoothed image (with a minimum significance of $3\sigma$) 
superimposed onto the binned image ([0.5-2] keV energy band). Contours are in steps of 0.2 counts
per 5.53" pixel, starting from 0.3. The outer region
denotes the locus where exposure time is 50\% of the on-axis exposure
time. 
}
\label{CL2015Xray}
\end{figure}

\subsection{X-ray data reduction} 

We have re-observed CL2015 with the X-ray telescope (XRT) on board the Swift satellite for 
a total of 67 ks.
We chose Swift because its low and stable background (Moretti et al. 2009) 
makes it the best choice for sampling a cluster expected to have low surface brightness 
(Andreon \& Moretti 2011). Indeed, XRT is 1.5 times more sensitive than X-ray Multi-Mirror Mission (XMM-Newton) 
to low surface brightness emission (Mushotzsky
et al. 2019; Walker et al. 2019) for the same exposure time.

For the Swift XRT data reduction and analysis we followed the procedures described in Moretti et al. 2009 for
extended sources, with the following improvements.
First of all, we were able to recover a section of the field of view previously discarded because
it was contaminated by calibration radioactive sources onboard Swift that are significantly decayed by now.
This allows us to recover $\sim$ 23\% of the area in the single visit image, and up
to $\sim60$\% when images with different orientations are combined 
together. Second,
measurements of extended sources of low intensity contaminated by a background require an accurate
knowledge of the background itself. Due to the low telescope orbit, the Swift background is lower and more stable
than XMM-Newton or Chandra; nevertheless, it displays very short periods of enhanced intensity and/or a spatial structure
due to increased CCD temperature and bright Earth light reflection. In order to filter out the
former, we applied a stricter filter to CCD temperature, $T < 55$ C. To eliminate low energy flares 
(due to the bright Earth light reflection),
we extracted the light curve in the [0.3-0.5] keV band from the level 2 event file in CCD external regions
(77000 pixel in total) and filtered out time intervals where the count rate exceeded
5 c/s.  Based on a more extensive 
XRT program led by us, we found that
by flagging periods of enhanced intensity at these very low energies, the background is lowered 
by a factor of up to three at low energies, with a loss of  
less than 5-10\% of the total exposure time.

The net observing time was 58 ks
after cleaning the data from increased background periods. 
Since the field of view of the observation is entirely filled with the cluster emission, we also use four 
fields (at high latitude) centered on gamma ray bursts to account for background. We removed 
the first part of the exposure, contaminated by the gamma ray burst, and reduced the
exposure time to resolve the background sources
at similar levels in the cluster and control field directions.  The final background exposures 
are 25, 37, 29, and 27 ks, which are lower than the cluster's because source detectability is 
easier in observations lacking a cluster diffuse emission.

Point sources are detected by 
a wavelet detection algorithm, and we masked pixels affected by them
when calculating radial profiles and fluxes.
In our analysis, we used energy-dependent exposure maps to calculate the effective
exposure time, accounting for dithering, vignetting, CCD defects, gaps, and
excised regions. 
Since Swift observations are taken with different roll angles that result in different 
exposures at different off-axis angles, 
we only consider regions where the exposure time is larger than 50\% 
of the central value. To check the validity of our analysis relative to that of different authors, methods, and telescopes, 
we also applied the same procedure to Abell 2029, which was also observed by Swift for 34 ks and
is in the XMM Cluster Outskirts Project (X-COP) sample (Ghirardini et al. 2019). 

\section{Results}

\subsection{Projected quantities}

Figure~\ref{CL2015Xray} shows the [0.5-2] keV image and the region considered here, limited
by the line defining where the exposure time 
is 50\% of the central value (roughly a circle with a diameter of $\sim 22\arcmin).$  
There are about 5400 net photons in the [0.3-7]
keV band in the unflagged regions. The X-ray
image lacks any strong evidence of bimodality or other irregularities, and the cluster
looks radially symmetric. 
The cluster center is 
iteratively computed 
as the centroid of X-ray emission within the inner 20 kpc. It is 12 kpc north of the Brightest Cluster Galaxy (BCG)
well within its optical size.

\begin{figure}
\centerline{%
\includegraphics[width=7truecm,angle=-90]{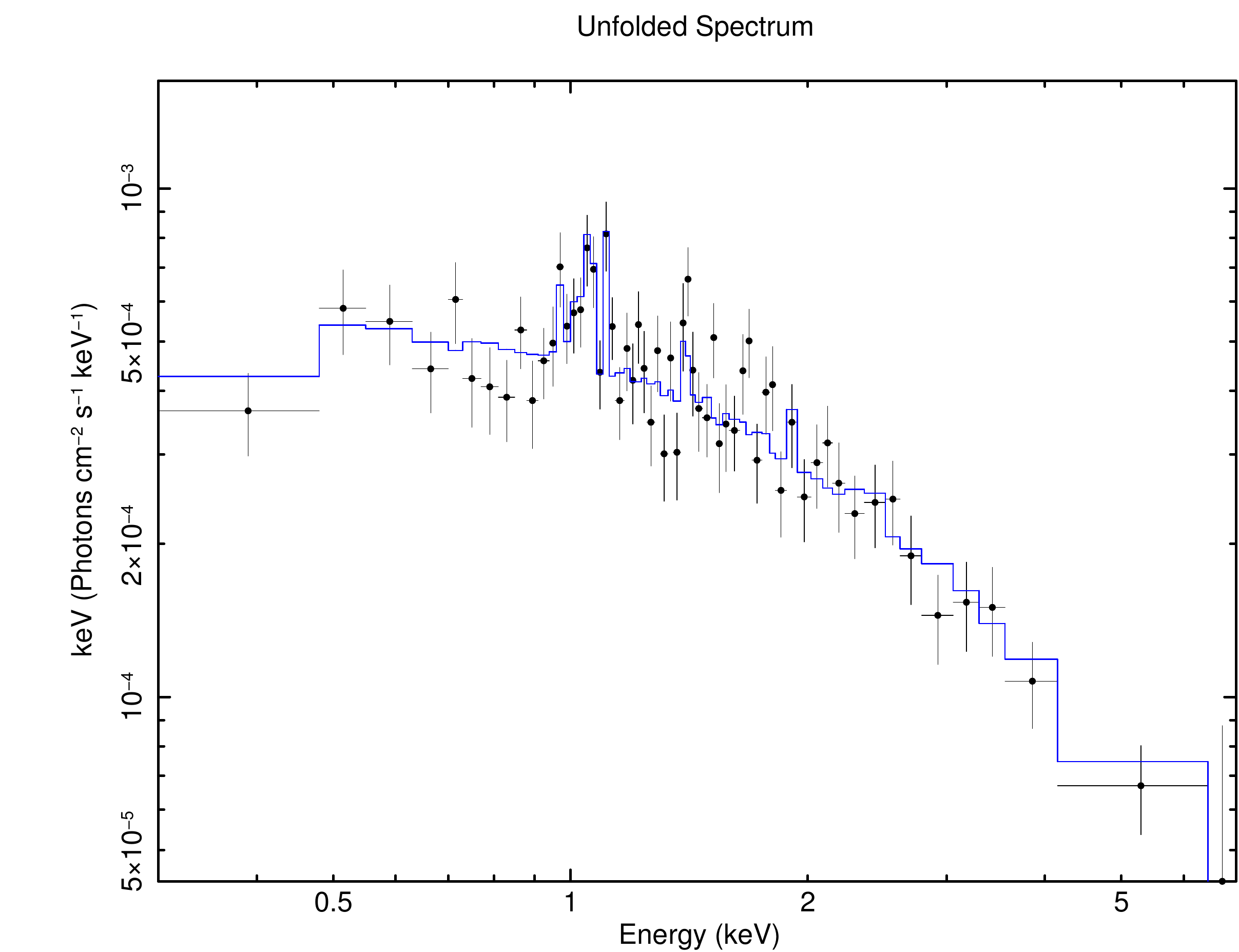}}
\caption{CL2015 X-ray spectrum in the $0.15 <r/ r_{500,cau}<0.5$ range 
and best fitting model. 
The spectrum is rebinned for display purposes, but is fitted
on a minimally binned version.}
\label{Xrayspec}
\end{figure}

We derived spectra in four different annular regions, and we assumed an absorbed 
Astrophysical Plasma Emission Code (APEC) model (Smith et al. 2001), with free metal abundance fixed at solar ratios (Anders \& Gravesse 1989).
We fixed the absorbing column at the Galactic value and   
the redshift at cluster redshift. The fit  
accounts for variations in exposure, flagged regions, and 
background, the latter being measured in the same angular annuli but in
our four background fields. 
The spectral counts were grouped to a minimum of five per
bin and  were fit with the 
XSPEC spectral package using the
modified C-statistic (also called W-statistic in XSPEC), in analogy with
Willis et al. (2005).
Simulations in Willis et al. (2005) 
confirm that resampling the data to prevent the occurrence of spectral bins
containing zero counts minimizes temperature biases
and this approach is currently used
by many authors.

\begin{figure}
\centerline{%
\includegraphics[width=6truecm]{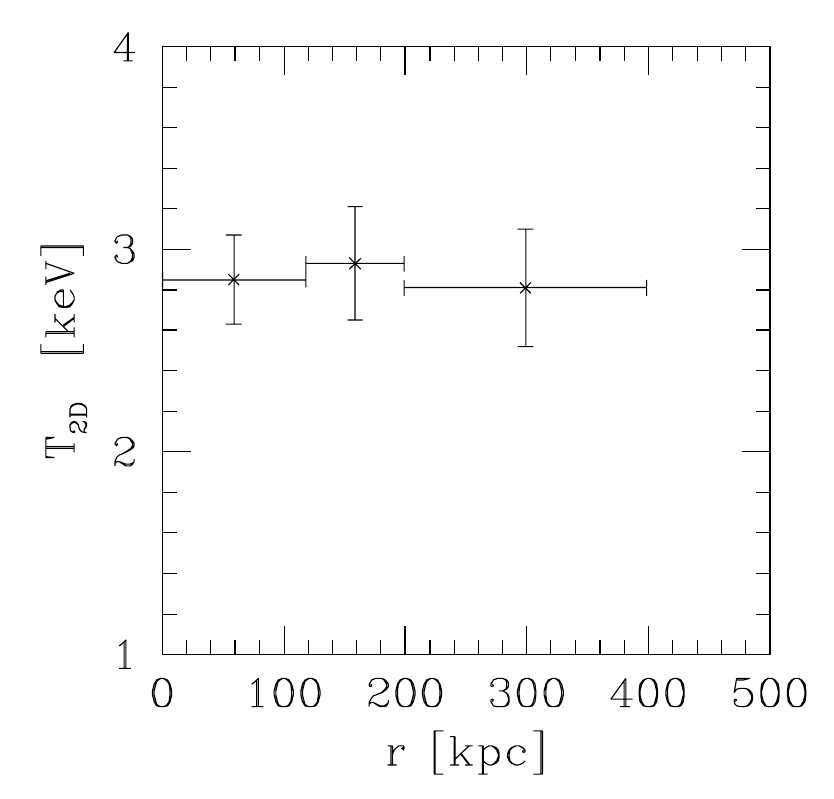}}
\caption{CL2015 projected temperature profile. 
}
\label{T2Dprof}
\end{figure}

Figure~\ref{Xrayspec} shows the spectral distribution of the counts 
(3400 net photons in $0.3-7.0$  keV band)
in the $0.15 <r/ r_{500,cau}<0.5$ corona, compared with the best fit model, at   
$kT=2.85\pm0.21$ keV and abundance $0.3\pm0.1$ solar. The line blend at $kT\sim1$ keV is
evident in the spectrum.
We also extracted spectra in the radial ranges  
$0.0 <r/ r_{500,cau}<0.15$, $0.15 <r/ r_{500,cau}<0.25$, and $0.25 <r/ r_{500,cau}<0.5$.   Each 
annulus contains $>1400$ net photons. We found a fairly flat projected temperature profile,
as shown by Fig.~\ref{T2Dprof}. Finally, we found no temperature differences among
the four quadrants with $1' <r <0.5 r_{500,cau}$.

We then measured the azimuthally-averaged surface brightness, which was corrected for exposure time,  vignetting, and
excluded regions, in the cluster and in the background fields. We measured the surface brightness 
in six energy bands ([0.3-0.5], [0.5-1], [1-2], [2-3], [3-5], and [5-7] keV) and 
in 21 annuli of increasing width to balance the decreasing signal-to-noise
ratio with increasing radius. The  smallest bin width allowed is 10$''$, comparable to the XRT point spread
function (PSF). 
Since the radial profiles derived in the four background fields are consistent, we used the 
averaged profile.

\begin{figure*}
\centerline{%
\includegraphics[width=14truecm]{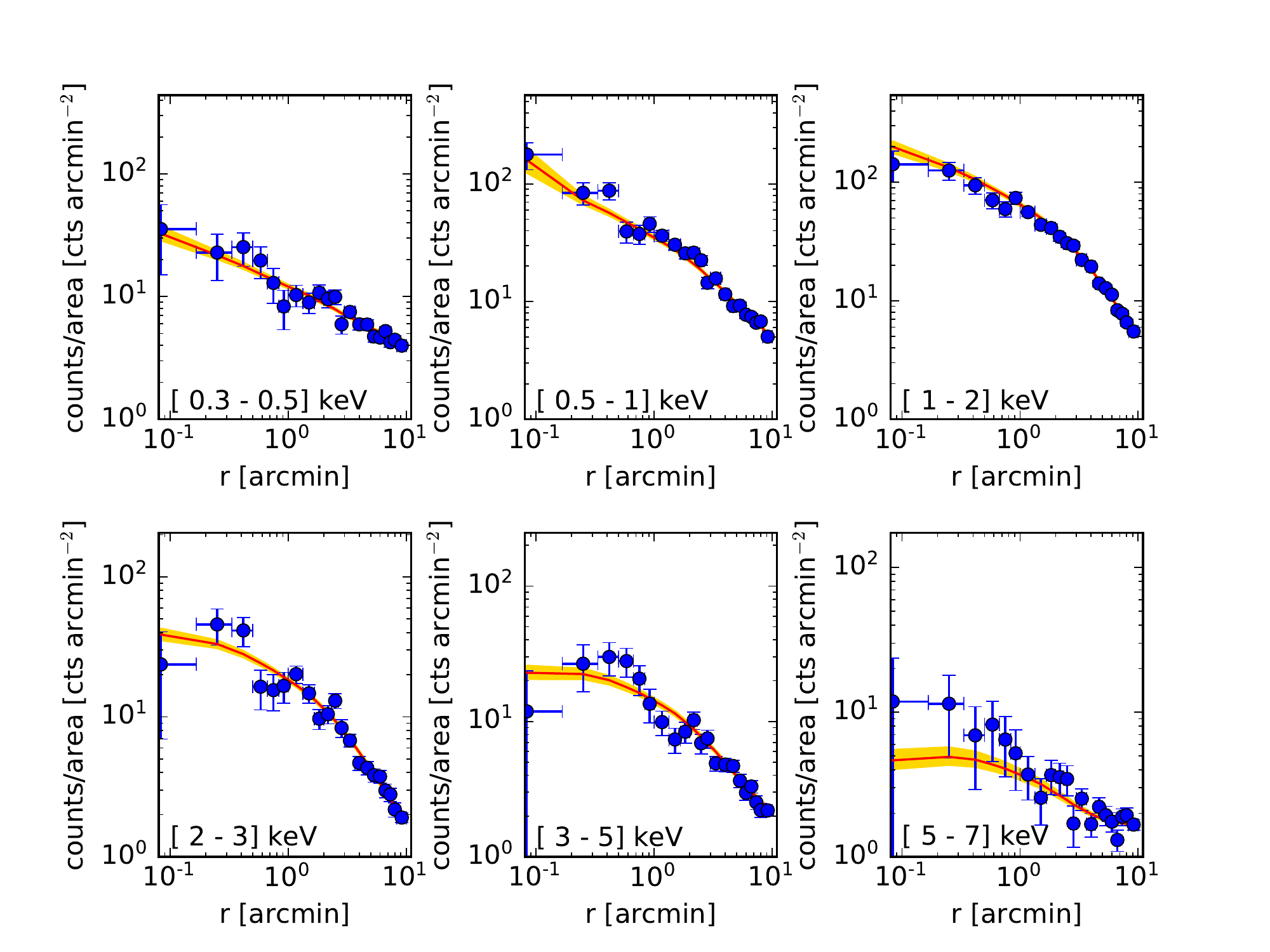}}
\caption{CL2015 radial surface brightness profiles.
Observed counts are indicated with points, whereas error bars mark $\sqrt{n}$ for display purposes
only.  The solid line is the mean fitted (physical) model (see Sect.~3.2) 
while the shading indicates the 68\% uncertainty.}
\label{CL2015counts}
\end{figure*}

Figure~\ref{CL2015counts} shows the distribution of the total counts per unit area in different 
bands (points).  
We have on average 45 net photons per energy-radius bin.

\subsection{Thermodynamic profiles}

Thermodynamic profiles were derived using MBPROJ2 (Sanders et al. 2018), a Bayesian
forward-modeling projection code that fits the surface brightness profiles (in different
energy bins) accounting for the background. The forward modeling goes beyond the limitations of previous
approaches, which were obliged to ``paste'' tailored-cut deconvolved and convolved profiles (e.g., 
Pratt \& Arnaud 2005),
apply arbitrary regularization kernels (e.g., Planck Collaboration 2013) 
to deal with noise, or ignore 
temperature gradients when deriving the electron density profile for clusters showing 
large temperature gradients (e.g., XCOP, Ghirardini et al. 2019).
As in other approaches, MBPROJ2 makes the usual assumptions about cluster 
sphericity, clumping, and, when requested, hydrostatic equilibrium.

We fit the data twice to understand the impact of assumptions on the derived
thermodynamic profiles.  In the first fit, a
descriptive analysis, we opted for an approach
similar to Vikhlinin et al. (2006) and we  modeled the temperature
and electron density profiles with flexible functions constrained by the data. This is a 
fit aimed at simply describing the observations, with models introduced to impose 
regularity and smoothness. 
In particular, the MBPROJ2 code models the electron
density as a modified single-$\beta$ profile following Vikhlinin et al. (2016):
\begin{equation}
  n_\mathrm{e}^2 = n_0^2
  \frac{(r/r_\mathrm{c})^{-\alpha}}{(1+r^2 / r_\mathrm{c}^2)^{3\beta-\alpha/2}}
  \frac{1}{(1 + r^\gamma / r_\mathrm{s}^\gamma)^{\epsilon / \gamma}}.\end{equation}

Similarly to Vikhlinin et al. (2016), the temperature profile is given by 
the product of a broken power law with three slopes and
a term introduced to model the temperature decline in the
core region:
\begin{equation}
\rm{T} = \textrm{T}_0\frac{((r/r'_c)^{a_{cool}}+(\textrm{T}_{min}/\textrm{T}_0))}
         {(1+(r/r'_c)^{a_{cool}})}\frac{(r/r_t)^{-a}}{(1+(r/r_t)^b)^{c/b}}.
\end{equation}
The other
thermodynamic profiles are derived from the ideal gas law. 
Following Vikhlinin et al. (2006), we fix $\gamma=3,$ and following Sanders et al. (2018) 
we use weak priors for the remaining six free electron density parameters.
Following McDonald 
et al. (2014), we
fix the inner slope to $a=0$ and the shape parameter of the inner region to $a_{cool}=2$.

In the second fit, we adopt instead a physical model for the cluster: we assume
a Navarro, Frenk, \& White (1997, NFW)
mass profile (for the dark matter only) and  hydrostatic equilibrium, which makes explicit temperature profile modeling unnecessary. 
In this case, MBPROJ2 computes
the pressure profile given the NFW mass profile 
and derives the other thermodynamic profiles
combining it with the electron density profile. As in our descriptive analysis, 
parameters are determined by fitting the data. 

In both fits, metallicity is a free parameter, 
absorption was fixed at the Galactic $N_H$ value in the direction of the cluster
from Kalberla et al. (2005),
and the results are marginalized over a further background scaling parameter to account
for systematics (differences in background level between the cluster and
control fields). 
The model is integrated on the same energy 
and radial bins as the observations, so that the results do not depend on
binning. 

In practice, the electron density profile is robustly determined
because it is the deprojected surface brightness profile (after a change of units 
with a minimal temperature dependence that is ignored in
major observational programs such as X-COP and REXCESS). 
Temperature is measured from the ratio 
of the cluster brightness in different energy bands, and temperature gradients are 
derived from radial variations in these
ratios. Radial temperature gradients are usually small, except in cool cores, 
to such a point that projected temperatures
are often overplotted on deprojected temperatures (e.g., Viklinin et al. 2005, Sun et al. 2009), which 
makes deprojection robust provided the radial profiles are regularized (by assuming a smooth
temperature or mass profile shape).
The other profiles come from $P=n_e T$ and $K= T / n^{2/3}_e$.

As done in the literature, the three-dimensional thermodynamic profiles are 
derived from the one-dimensional profiles (cf.
in Fig.~\ref{CL2015counts}) under several assumptions that include: a) spherical
symmetry, 
that is, concentric isodensities of identical center, position angle, and zero ellipticity; b) smooth,
unclumped gas distribution without any substructure; and c) 
how and how much the observed 
radial profiles are regularized or the fitted
model is constrained by assumptions. 
In particular, the errors computed (by us and others) depend on all these
assumptions and, given that clusters are known to deviate somewhat from the ideal 
assumed behavior, 
errors should always be regarded as an underestimate
of the true uncertainty. 

\begin{figure}
\centerline{\includegraphics[width=9truecm]{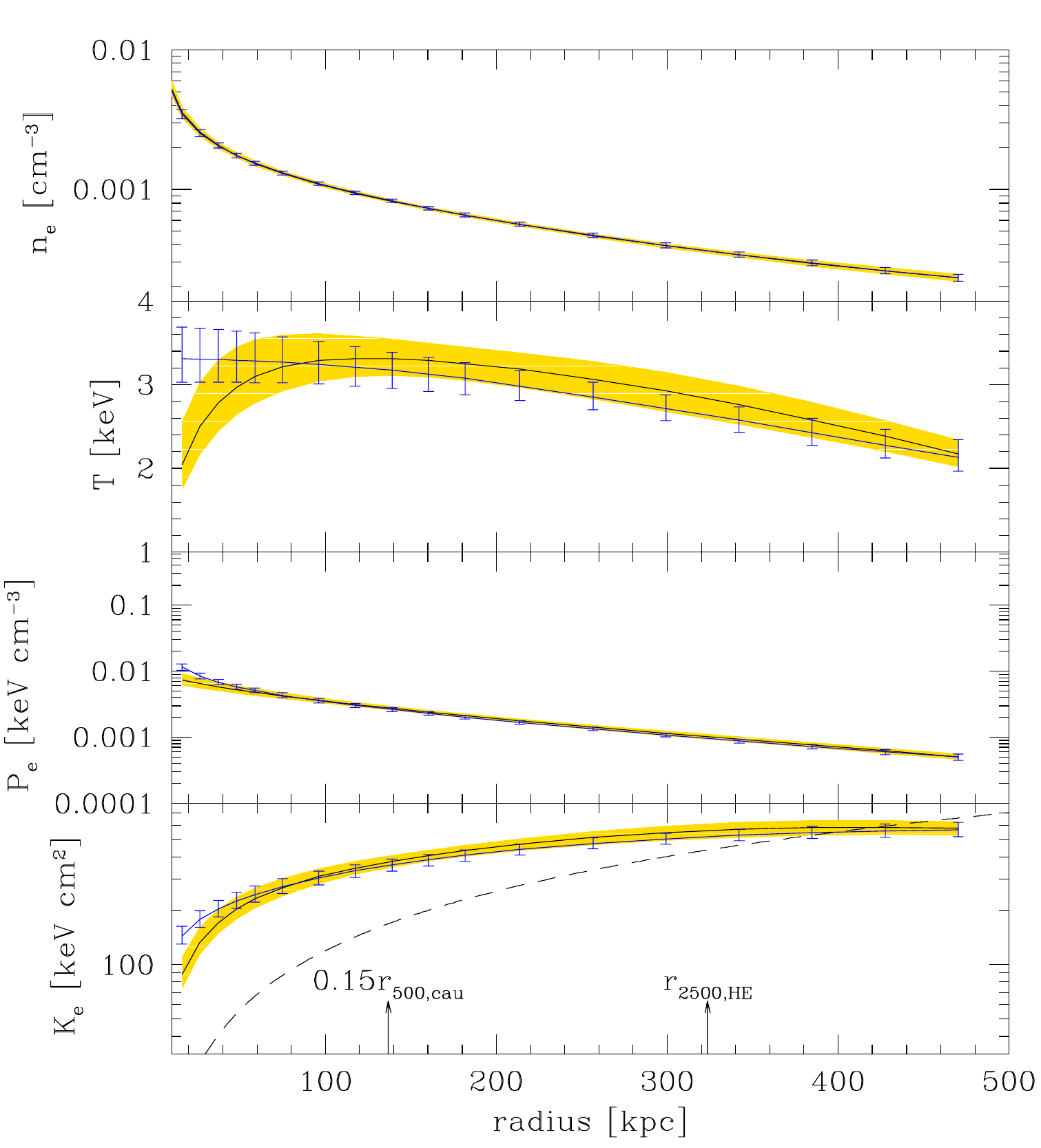}}
\caption{CL2015 thermodynamic profiles. The solid line with shading shows
our physical fit, while the solid line with error bar is our descriptive fit.
In both cases we plot the mean model and 68\% errors. The dashed curve in the entropy
panel is the Voit et al. (2005) fit to non-radiative simulations, as adapted by Pratt et al. (2010).
}
\label{CL2015phys_twometh}
\end{figure}

\begin{figure}
\centerline{\includegraphics[width=9truecm]{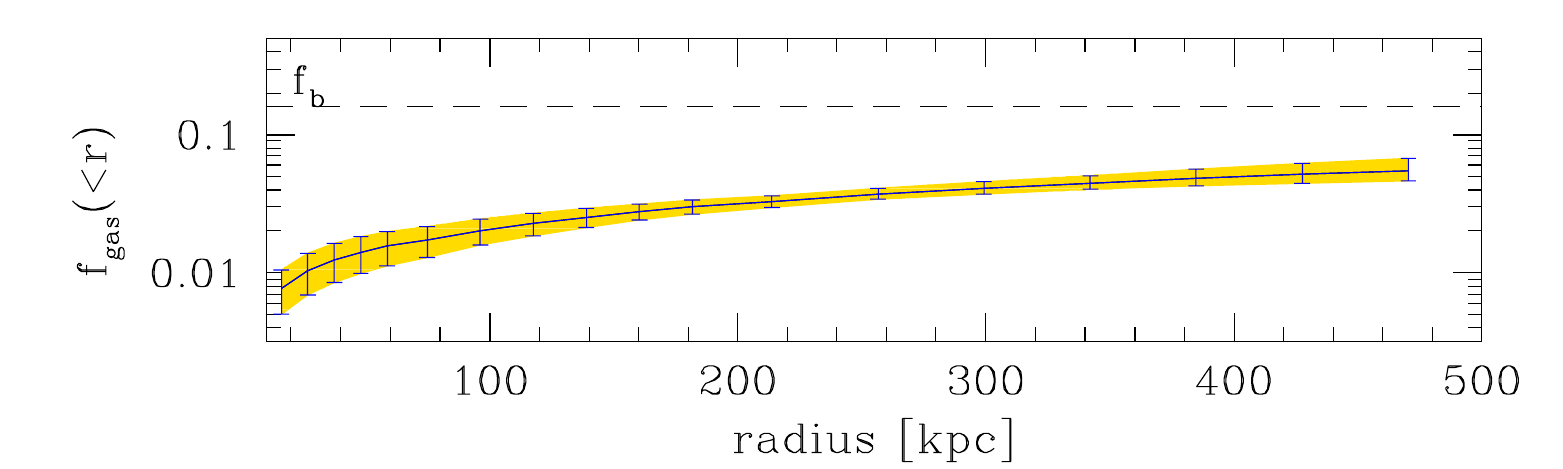}}
\caption{CL2015 enclosed gas fraction profile from the physical modeling. The horizontal dashed line shows
the Universe value.}
\label{CL2015fgas}
\end{figure}

To test our ability to derive thermodynamic profiles against state-of-the-art analyses, we compared the 
thermodynamic and mass profiles of Abell 2029 derived by us using Swift with the same technique
described here for CL2015 with those derived
using XMM-Newton by Ghirardini et al. (2019). The
profiles agree well with each other, as detailed
in the Appendix.

CL2015 surface brightness profiles are well fitted in the various bands by our physical model  
(Fig.~\ref{CL2015counts}). This is also true for the descriptive  model.
Profiles extend to about $r_{1000}$ ($r_\Delta$ values are derived in the next section), 
where the cluster contribution is still $>5$\% of the background level.  
The model captures well the data trend, being
almost always within (approximately) $1\sigma$ of the data. This is unsurprising given that we used 
flexible functions, with 11 or 14 free parameters, that act as regularization kernels, to model the data.

Figure~\ref{CL2015phys_twometh} shows CL2015 thermodynamic profiles derived both in our
descriptive analysis (solid line with error bars) and with the assumption of hydrostatic equilibrium and a NFW 
mass profile (solid line with shading). 
The descriptive and physical profiles, which are posterior distributions, 
turn out to be indistinguishable 
where the data (the likelihood) dominate, while they differ in the low signal-to-noise (S/N) regime, where the prior
dominates, namely
at the very center, for three out of the four thermodynamic quantities considered, 
as also discussed in Sanders et al. (2018) for a different sample.
Differences at the center are due to the small volume within a sphere of small radius $r$, and
to the presence of many photons emitted outside the sphere but inside 
the cylinder of radius $r$, which lowers the S/N of the
central quantities. Since the signal coming from the central volume is noisy, the prior
dominates. Instead, quantities measured outside the very center and 
global quantities, such as the mean or integrated value within $r_{2500}$, are not affected by these 
modeling differences. In summary, derived thermodynamic profiles
are robust to (tested) assumptions, with the exception of the very center for strongly $T$-sensitive
quantities. We tested that the electron density profile derived here agrees with our previous 
determination (Andreon et al. 2017a) from shallower Swift data, which required stronger assumptions.

We found  $\log K_{2500} = 2.71\pm0.03$ keV cm$^{-2}$, 
$T_{2500}=2.6\pm0.2$ keV, and $f_{gas,2500}=0.043\pm0.005$, almost
indepedent of the fitting model (see Fig.~\ref{CL2015phys_twometh}), and an abundance
$0.23\pm0.12$ solar  
for both fits (the latter in agreement with the 2D analysis).
Figure~\ref{CL2015phys_twometh} shows that in CL2015 
there is an entropy excess compared to the non-radiative simulations
of Voit et al. (2005), as commonly found in clusters.
Figure~\ref{CL2015fgas} shows the enclosed gas fraction profile derived with the physical
fit.

\begin{figure}
\centerline{%
\includegraphics[width=9truecm]{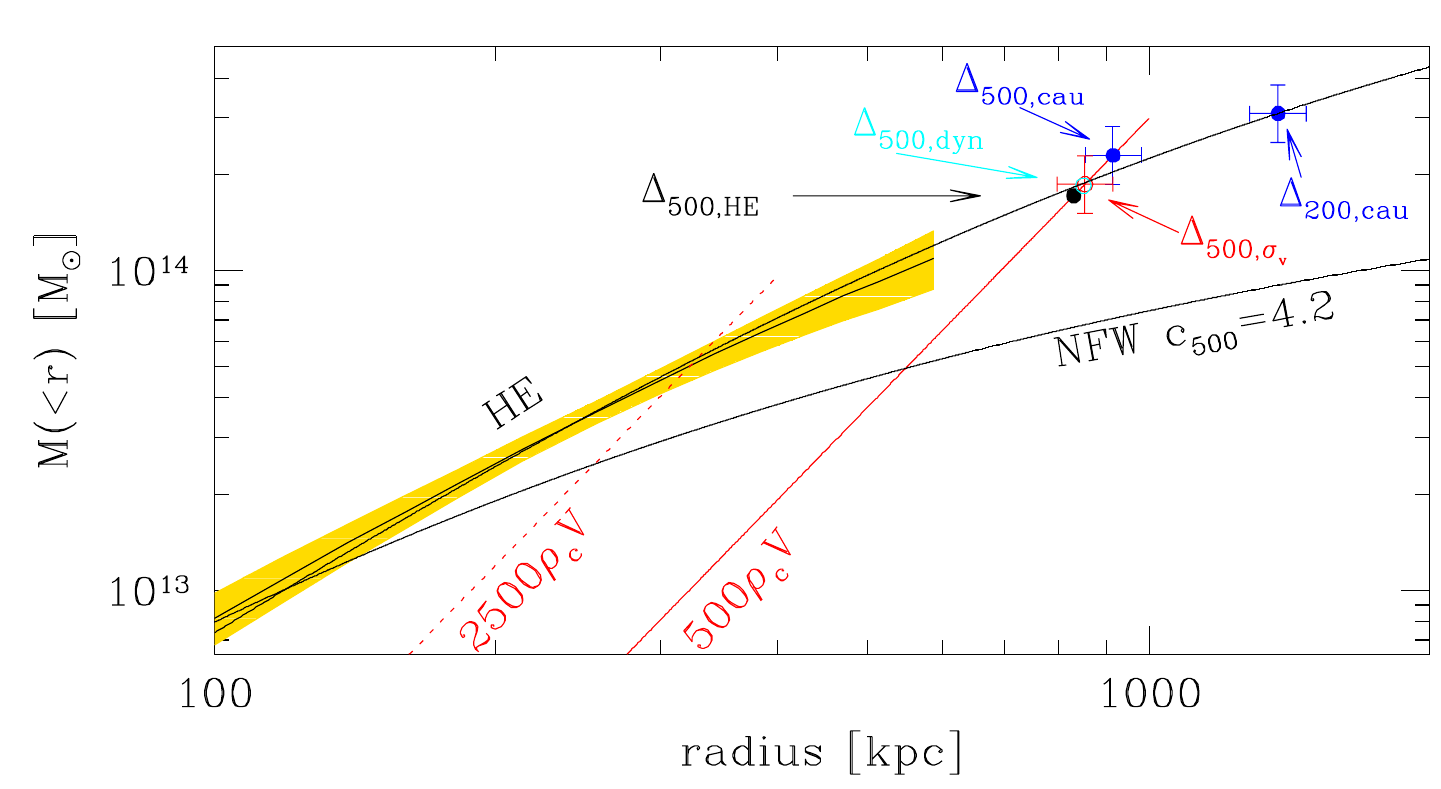}}
\caption{CL2015 mass comparison. Hydrostatic mass (HE) within the radius $r$ (black
solid curve) is compared to $500\rho_c V$ (solid red line), 
$2500\rho_c V$ (dotted red line), and
to caustic, velocity-dispersion-based, and other dynamical masses (blue, red, and cyan points, respectively).
We also plot a NFW profile with $c_{500}=4.2$ normalized at $r=100$ kpc (bottom curve)
and NFW profile with $c_{500}=1.5$ normalized at $r_{200}$ (top curve fitting all data).}
\label{McompCL2015}
\end{figure}

\subsection{Concentration of the cluster mass}

Figure~\ref{McompCL2015} plots the mass profile $M(<r)$ fitted to our X-ray
data under the assumption of hydrostatic equilibrium. The intercept with $\Delta \rho_c V$
gives, by definition, $r_{\Delta,HE}$ and $M_{\Delta,HE}$, where the subscript ``HE" emphasizes
that we assume hydrostatic equilibrium. Our data reach $\Delta=1000$.
This is the same radius covered with the sample of clusters in 
Arnaud, Pointecouteau, \& Pratt (2005), which is used to 
calibrate the REXCESS (e.g., Pratt et al. 2009) and Planck (e.g., Planck Collaboration 2014) 
scaling relations.
Values at overdensity $\Delta=2500$ are well measured, with no extrapolation, and we find 
$\log M_{2500,HE}/M_\odot = 13.69\pm 0.09$ (and therefore $\log r_{2500,HE} = 2.51 \pm 0.03 $).
Values at $\Delta=500$ require some extrapolation, and
we find $\log M_{500,HE}/M_\odot = 14.23\pm 0.22$ (and therefore $\log r_{500,HE} = 2.92 \pm 0.08$).
The extrapolation to $\Delta=500$ is commonly done even for clusters with very high quality 
data, such as those used by Arnaud et al. (2005).

\begin{figure}
\centerline{\includegraphics[width=7truecm]{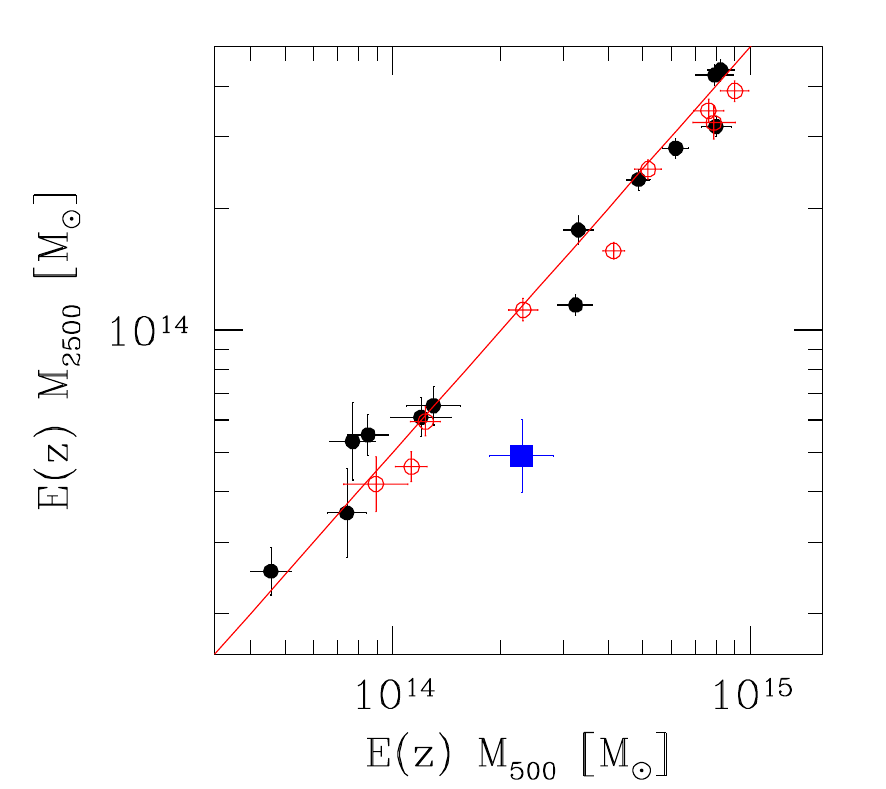}}
\caption{Mass at different overdensities comparison. Masses at two different overdensities, $\Delta=500$ and $\Delta=2500$, of 
CL2015 (blue point), and clusters in Vikhlinin et al. (2006) and Sun et al. (2009) (solid points) and 
Arnaud et al. (2005, open points) are shown. The red line is the 
location $M_{500}/M_{2500}=2.0$.}
\label{M2500M500}
\end{figure}

\begin{figure}
\centerline{\includegraphics[width=7truecm]{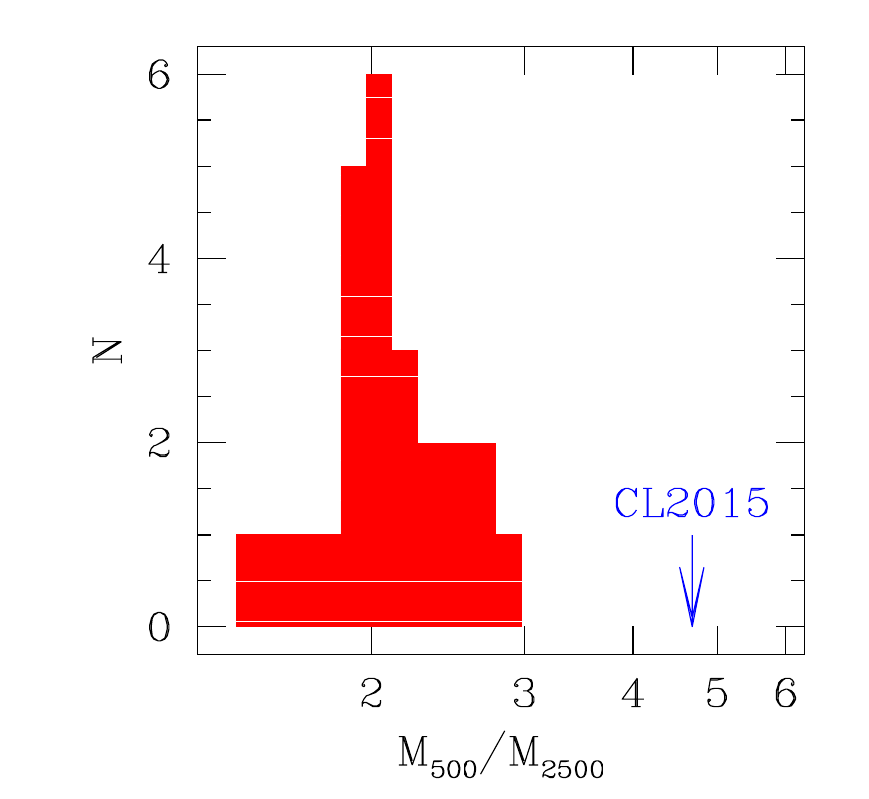}}
\caption{Inverse sparsity comparison. The arrow marks the value of
$M_{500}/M_{2500}$ of CL2015 while the histogram shows the distribution of
clusters  in Vikhlinin et al. (2006), Sun et al. (2009), and 
Arnaud et al. (2005). }
\label{histosparsity}
\end{figure}

\begin{figure}
\centerline{\includegraphics[width=7truecm]{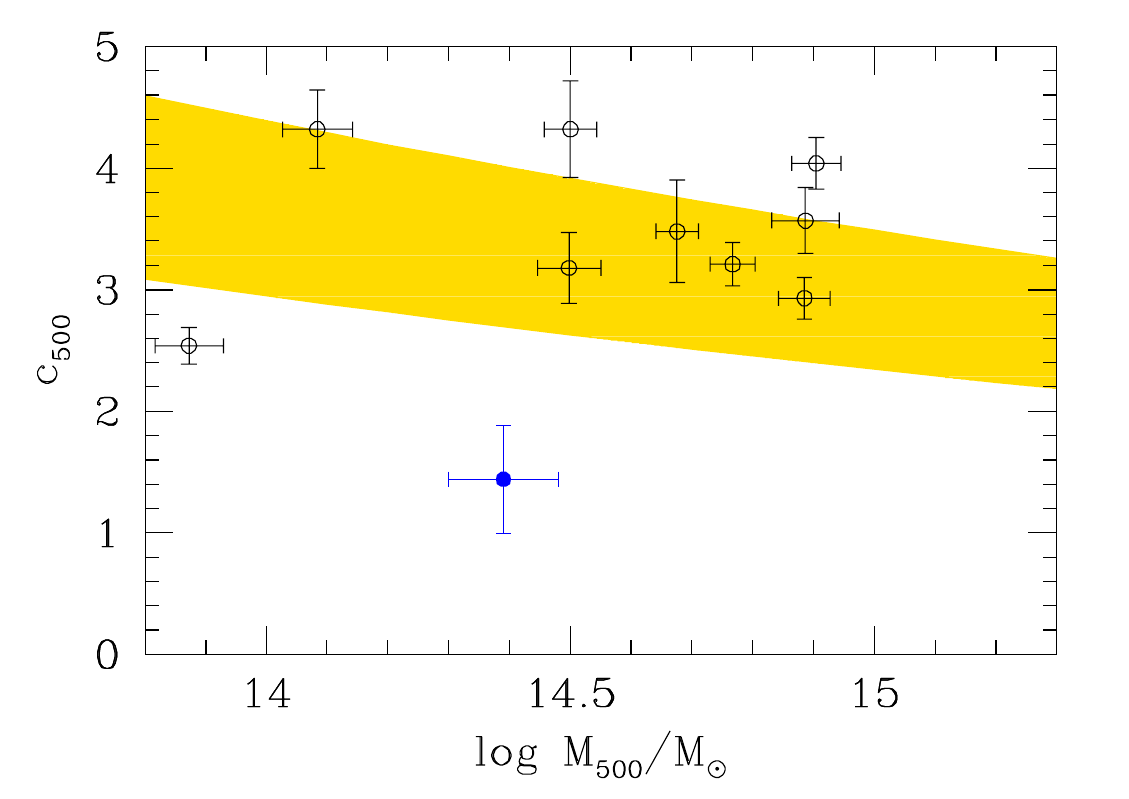}}
\centerline{\includegraphics[width=7truecm]{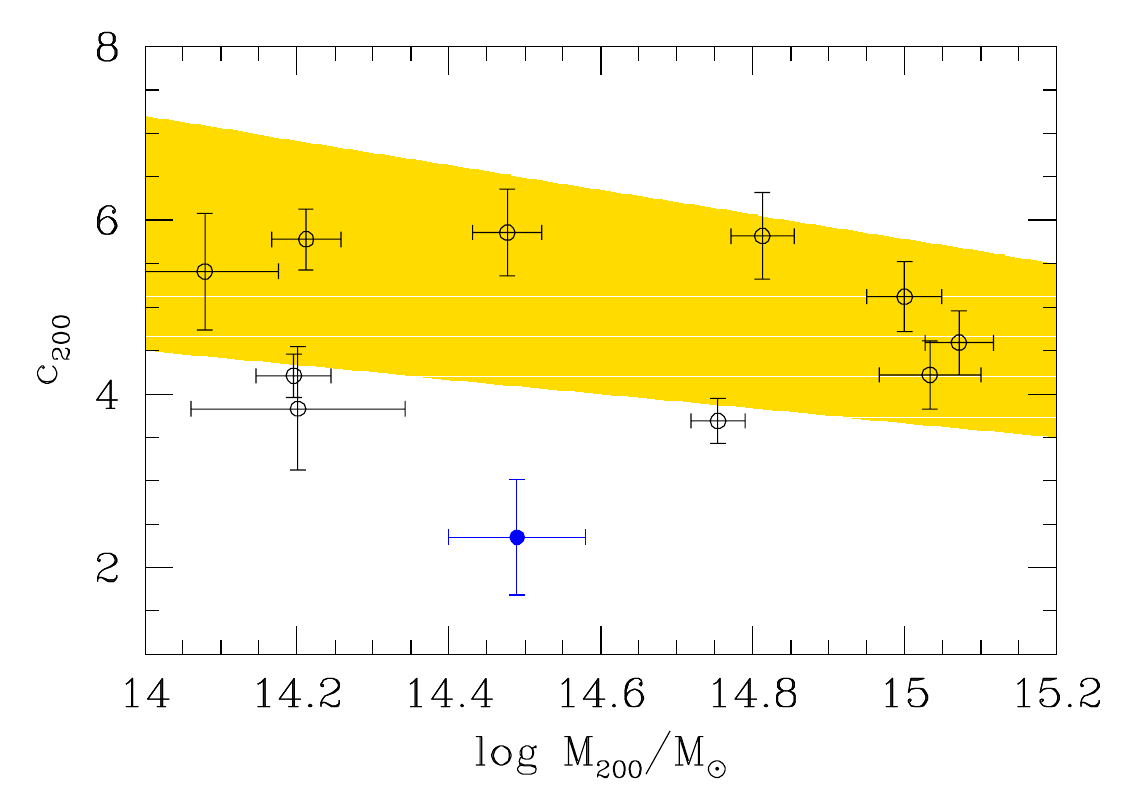}}
\caption{Concentration comparison. Top and bottom panels compare CL2015 to clusters in Vikhlinin et al.
(2006) and to clusters in Pointecouteau et al. (2005). The shadings indicate $\pm1\sigma$ scatter
around the theoretical prediction by Dolag et al. (2004).}
\label{conc}
\end{figure}

Figure~\ref{McompCL2015} also shows the other estimates of mass of CL2015 at $\Delta=500$, from the three
dynamical analyses by two different teams (Sect~2.1 and Appendix~A). All four estimates of mass, 
of which one is hydrostatic
and three are dynamical, agree with each other.  No matter which 
$M_{500}$ is taken, CL2015  remains
an outlier, at $>5\sigma$, in the REXCESS $L_X-M_{500}$ relation.

Figure~\ref{McompCL2015} also shows the estimate of caustic mass at the best constrained overdensity, 
$\Delta=200$ ($M_{200}=14.49\pm0.09$), and
a NFW profile with $c_{500}=4.2$, a typical value measured
in hydrostatic analyses (e.g., Vikhlinin et al. 2005, Sun et al. 2009), 
arbitrarily normalized at $r=100$ kpc.   
A NFW profile with such a high concentration, however, does not reproduce the mass distribution of CL2015: 
with the adopted normalization, there is a significant mass excess at large radii in CL2015, or
deficit at small radii, relative to other clusters with $c_{500}\sim4.2$. 
This is true also ignoring dynamical estimates of mass.
A lower value of $c_{500}=1.5\pm0.4$  instead fits all data (upper curve in Fig.~\ref{McompCL2015}).

Figure~\ref{M2500M500} reiterates the point: clusters with a given value of the ratio
$M_{500}/M_{2500}$ (inverse sparsity in Corasaniti et al. 2018), would be aligned 
on a line parallel to the one drawn, $M_{500}/M_{2500}=2.0$, 
appropriate for a cluster of $c_{500}=4.2$.  This ratio is observed in the samples of 
(morphologically relaxed) clusters in Vikhlinin et al. (2006), Sun et al. (2009, closed points),
and Arnaud et al. (2005, open points). In the figure, to filter out noisy measurements 
we only consider clusters with measured
$M_{500}$ with errors less than 0.1 dex and we
removed clusters noted by the authors as having largely extrapolated $r_{500}$ or 
values that should be treated with caution. CL2015 is a clear outlier 
in concentration, with $M_{500}/M_{2500}=4.7$.
Figure~\ref{histosparsity} reinforces this point: with $M_{500}/M_{2500}=4.7$, CL2015 is extreme when 
compared to values in the above cluster samples. 
Since we have centered the cluster on the X-ray peak (and the BCG), 
the observed low concentration does not come from having mis-centered the
cluster.
CL2015 would still be an outlier in sparsity if we used its hydrostatic mass instead of caustic mass, 
which would give $M_{500}/M_{2500}=3.8$. 

From masses at $\Delta=2500$ and $\Delta=200$ we derived a concentration of $c_{200}=2.4\pm0.7$ and 
$c_{500}=1.5\pm0.4$.  Figure~\ref{conc} compares CL2015's concentration
with determinations from Vikhlinin et al. (2006) and Pointecouteau et al. (2005), and
theoretical predictions from Dolag et al. (2004).  In the sample of Pointecouteau et al. (2005), all
measurements are extrapolated to $\Delta=200$, since, as mentioned, these reach 
$\Delta\sim1000$ only. Also in these plots, CL2015's concentration is low, and also 
about $2\sigma$ below the theoretical predictions of Dolag et al. (2004), a value still in
the range of values that we expect to see in samples.

\begin{figure}
\centerline{%
\includegraphics[width=6truecm]{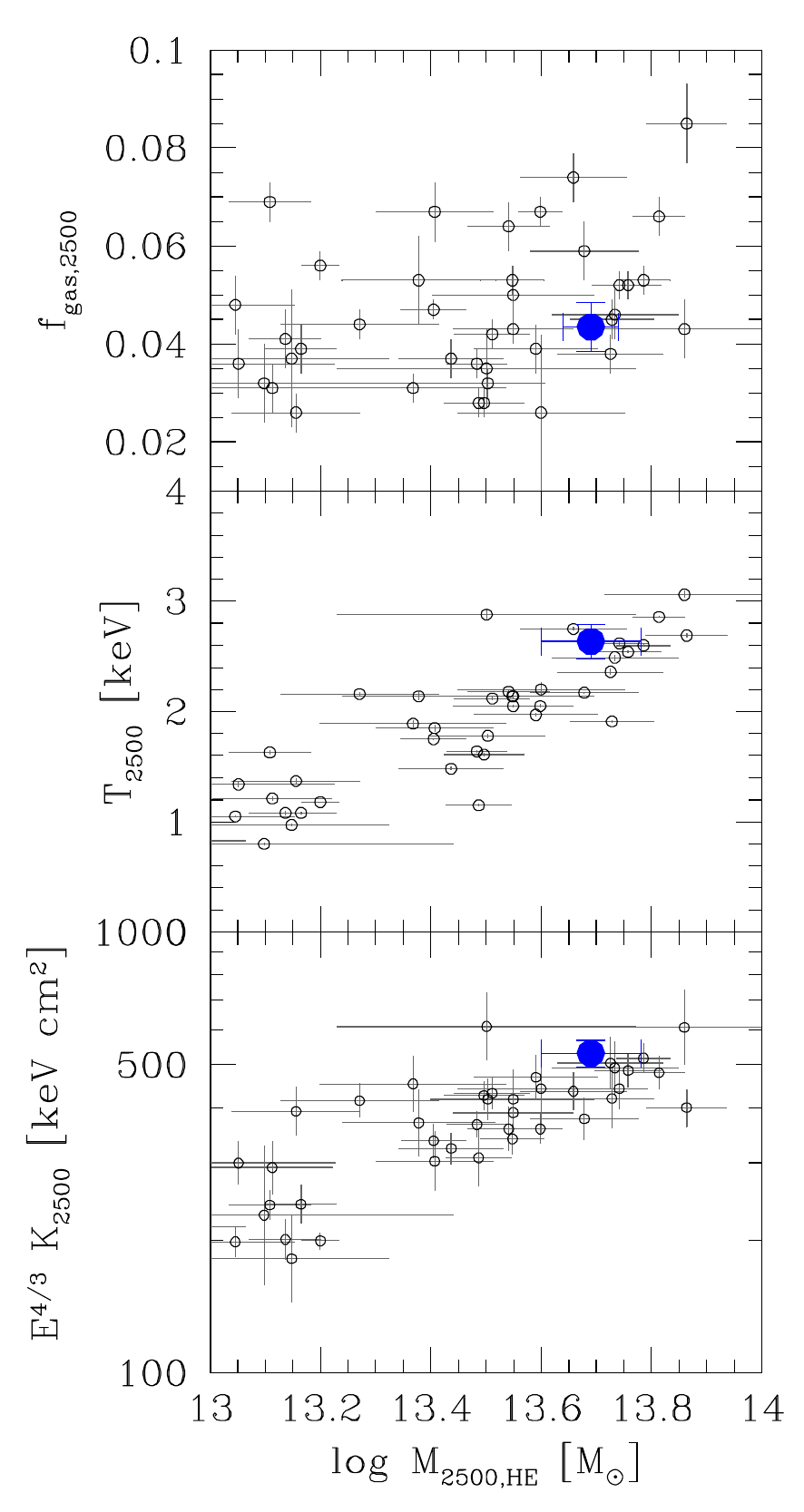}}
\caption{Enclosed gas fraction (top panel), temperature (middle panel), and entropy (bottom panel) at
$r_{2500,HE}$ of CL2015 (solid blue point) and of
relaxed groups and clusters in Sun et al. (2009, open points).}
\label{CL2015vsSun}
\end{figure}

\subsection{What causes the low X-ray luminosity of CL2015?}

We now show that, when compared to clusters of the same mass at the overdensity $\Delta =2500$, 
CL2015 is a normal object. Furthermore, we show that the low X-ray luminosity of CL2015 is due to the low
total mass in the inner part of the cluster compared to other clusters of the same mass at 
$\Delta =500$ and to the different dependence of mass
and luminosity with radius (most of the mass within $r_{500}$ is close to $r_{500}$ while
most of the X-ray photons come from $r\ll r_{500}$).

To show this, we compare CL2015 to the 43 groups and
clusters  in Sun et al. (2009). Sun et al. (2009) estimate masses assuming 
hydrostatic equilibrium, which limits them to consider only   
morphologically relaxed clusters. 
Figure~\ref{CL2015vsSun} shows the distribution of entropy, temperature, and enclosed gas fraction
for the cluster sample in Sun et al. (2009) as a function of $M_{2500,HE}$. The position of 
CL2015 in these plots is entirely consistent with the other clusters.
In other words, CL2015 is not gas poor, or an outlier, relative to other clusters of similar mass at 
overdensity $\Delta=2500$.

CL2015 becomes an outlier when its different concentration plays a role. This is because
other clusters with the same mass at $\Delta=500$  
have a much larger $M_{2500}$. For a normal
gas fraction, the lower $M_{2500}$ results in a low gas mass and a low X-ray luminosity. 
With all other properties fixed,  a deficit of \textasciitilde$5.5$ ($=(4.7/2)^2$) in $L_X$ is expected, 
which is comparable to what is observed (Fig.~\ref{selLx}). A better estimate of this deficit 
would require
taking into account the concentration at all radii, rather than at $r_{2500}$ only.

We conclude that CL2015 stands out in the $L_{X,500,ce}$ versus $M_{500}$ relation because
although these quantities are integrated on the same radial range, this does not suffice
to remove the difference dependency with radius of 
luminosity and mass: most of the X-ray photons are
emitted from regions at small radii, 
while most of the mass is at larger radii. Integrating within the same radial
range, $0.15< r/r_{500}<1$, does not completely remove these dependencies.
Once concentration is factored out by considering clusters
of the same mass at the radius $r_{2500}$ at which differential quantities  
are measured, CL2015 falls within the distribution in 
temperature and entropy occupied by the Sun et al. (2009) clusters (Fig.~\ref{CL2015vsSun}). This also holds
true for enclosed gas density because most of the gas mass is close to the outer integration radius.

\begin{figure}
\centerline{\includegraphics[width=7truecm]{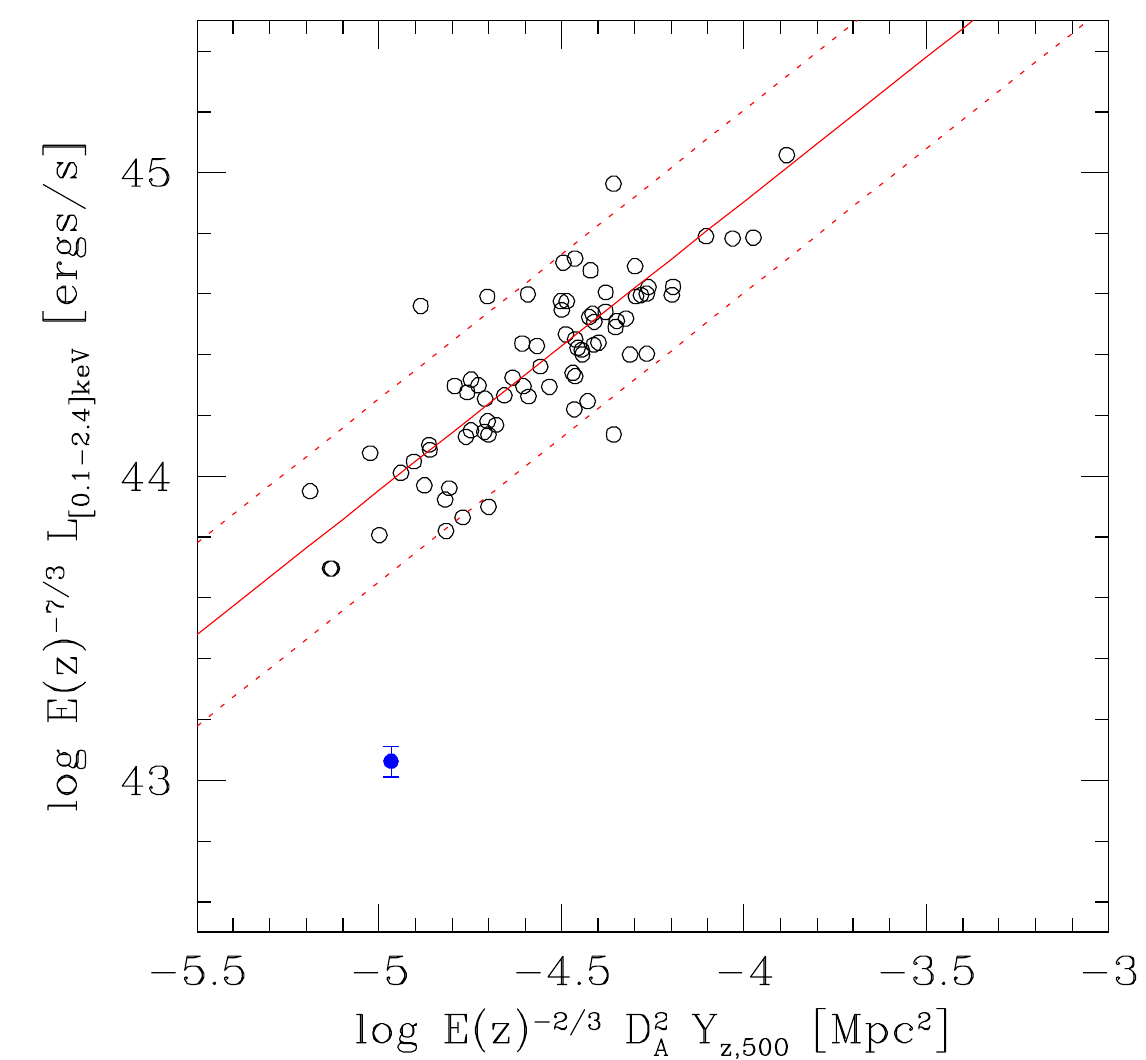}}
\caption{X-ray luminosity vs. integrated pressure parameter. The blue point is CL2015,
open points are $z<0.2$ clusters in both the REFLEX and Planck collaboration (2014) catalogs.
The dashed corridor indicates $\pm$ a factor of two.}
\label{LxYz}
\end{figure}

\subsection{Independent proof: The SZ view}

We can test whether our explanation of the outlier position of  CL2015 is valid 
by looking at scaling relations of the
X-ray luminosity with another quantity that has a different dependence with radius,
such as the integrated Compton parameter, that is, integrated pressure. 
Since pressure is proportional
to $n_e$, while L$_x \propto n^2_e$, the integrated pressure depends on 
the integrated gas mass, which is a steeply growing
function almost unaffected by a deficit of mass at the center, which instead strongly affects 
$L_X$, suggesting that 
CL2015 could be an outlier also in the $L_X-Y_{SZ}$ plane.

Figure~\ref{LxYz} shows the $L_X$ versus $Y_{z,500}$ for REFLEX clusters with $z<0.2$ 
(Bohringer et al. 2004) in the Planck Collaboration (2014) catalog (open points). The red solid 
line is the relation expected for REXCESS clusters, as calculated in Planck Collaboration (2011).
CL2015 is detected by Planck (Planck Collaboration 2014)
and is also plotted in the figure (blue solid point) with the X-ray luminosity converted from
the [0.5-2] keV Swift value. CL2015 is an obvious outlier in the $L_X-Y_{500}$ plane.  
This was also noted in Planck Collaboration (2016), where it was considered   
``X-ray underluminous" for its SZ strength based on an X-ray luminosity (presumably un upper-limit,
not listed) from Rosat All Sky Survey data.

To summarize, CL2015 is an obvious outlier compared to easy-to-detect (REXCESS, REFLEX, Planck) clusters 
in scaling relations involving X-ray luminosity (measured
by three different teams from three different X-ray telescopes) versus dynamical mass (measured by two teams in
three different ways) or integrated pressure (measured by two teams). This reassures us that the outlier status
of CL2015 is not a telescope temporary failure or a momentary lapse of
reason on the part of the author. The low concentration is the ultimate reason for CL2015's outlier status 
in quantities that weight cluster-centric radii in different ways
(namely $L_{X,500,ce}-M_{500}$ and $L_{X}-Y_{z,500}$) and also its normal behaviour for
differential measurements performed at the same radius (namely, $T_{2500}-M_{2500}$, $K_{2500}-M_{2500}$) or close radii 
($f_{gas,2500}-M_{2500}$). 

We emphasize once more that CL2015, at 
$\sim -1.5\sigma$ from the relation in an X-ray unbiased sample, 
is not an outlier in that distribution. 
The rarity of outliers in 
$L_X-Y_{SZ}$ or $L_X-M_{500}$
when easy-to-detect cluster samples are considered is related to the difficulty of having 
such type of objects
in samples because of their low X-ray luminosity for their $M_{500}$ and also their low SZ signal.
CL2015 is at the very boundary of the Planck detection limit: it is detected at the threshold
of one of the detection codes ($S/N=4.57$, MMF1, Planck Collaboration 2016) 
and below the threshold of a different implementation of the matched multifilter code (MMF3).
No other XUCS clusters of low X-ray luminosity for their mass or
with $\log M_{gas,500}/M_\odot <13.1$ are detected by Planck.
Therefore, even Planck is missing clusters with $\log M_{500}/M_\odot \approx 14.3$ unless
they have a gas fraction larger than CL2015.

\begin{figure}
\centerline{\includegraphics[width=6truecm]{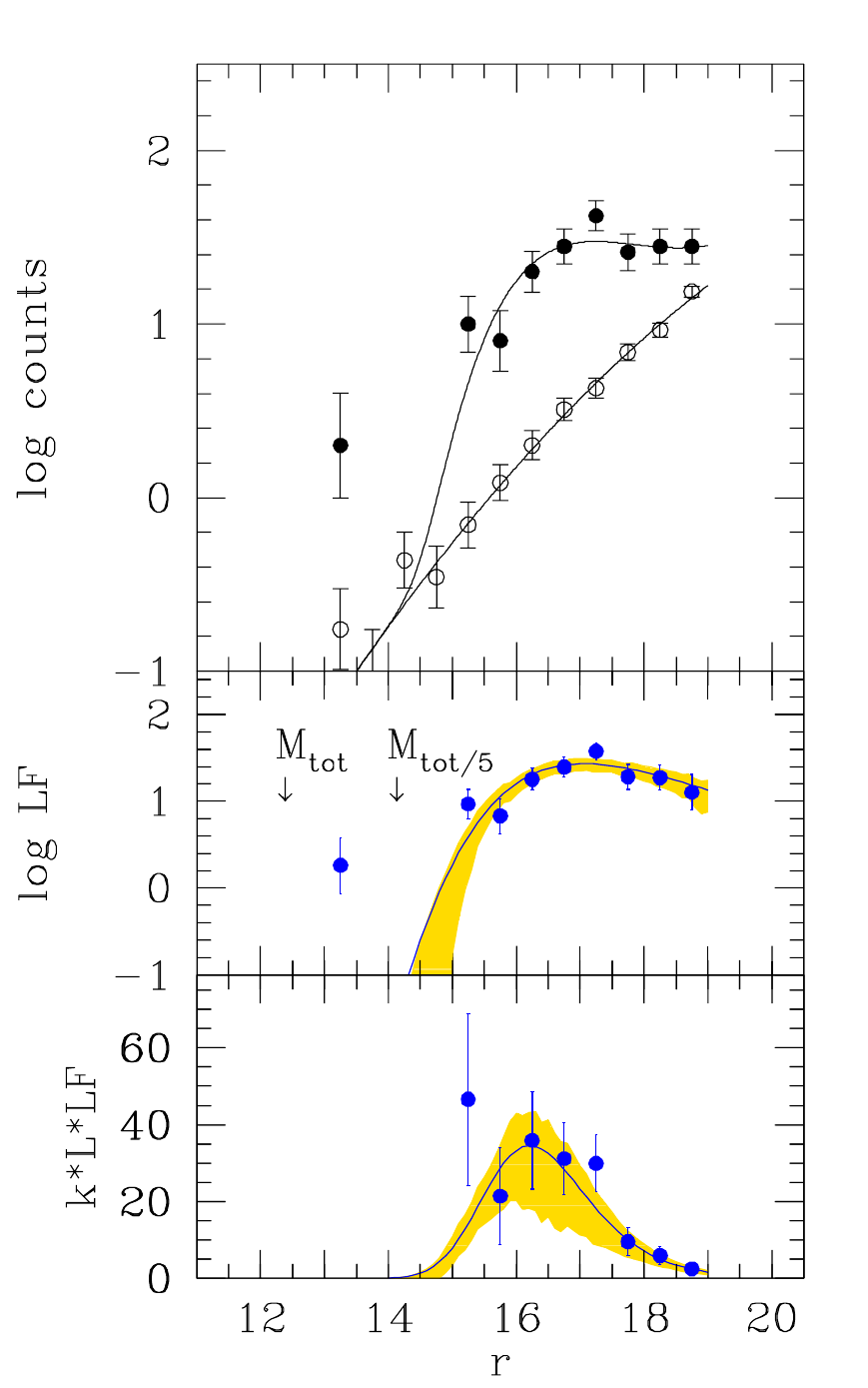}}
\caption[h]{Luminosity Function determination. The upper panel shows galaxy $\log$
counts in the CL2015 direction (solid dots),
in an adjacent line of sight for background estimation (open dots). The cluster contribution is given
by the difference of the two counts, shown in the middle panel, which
also reports the integral of the luminosity function ($M_{tot}$)
and $1/5^{th}$ of it ($M_{tot/5}$). The bottom panel
gives the luminosity contribution of each magnitude bin (in arbitrary units). 
Curves mark the fitted model to un-binned data, after removal of the brightest 
cluster galaxy.
Approximated point errors (computed with the usual sum in quadrature)
are marked with bars; precisely computed 68\% errors on the model are shaded. }
\label{LF}
\end{figure}

\begin{figure}
\centerline{\includegraphics[width=7truecm]{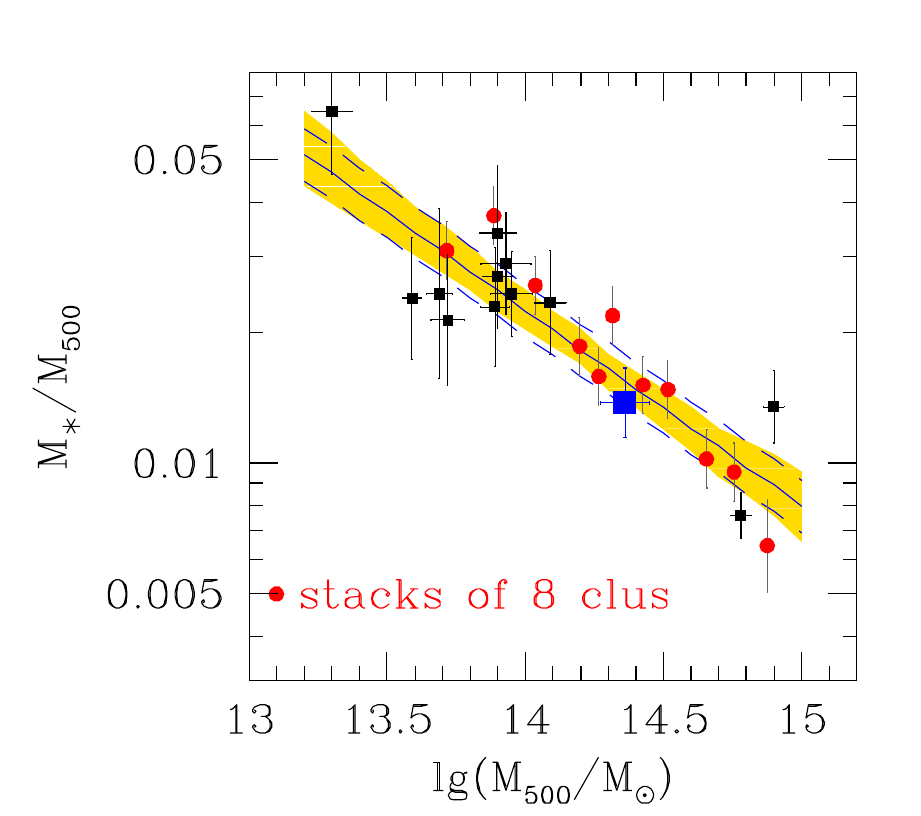}}
\caption{Stellar fraction of CL2015 (solid blue square) compared to other clusters: a) X-ray relaxed
clusters (black points) from Andreon (2012), with stellar mass of Abell 2029 updated in Renzini \& Andreon (2014).
The solid line is the mean model, 
while the shading indicates the 68\% uncertainty
(highest posterior density interval). The dashed corridor is the mean model $\pm$ the intrinsic
scatter. b) X-ray selected clusters (red points), after averaging in bins of eight clusters each to 
emphasize the mean trend while reducing the
scatter around the mean relation.}
\label{stellar_fraction}
\end{figure}

\subsection{One more test: Stellar content}

An alternative test to support our suggestion of a shallower mass distribution in CL2015 relative
to other clusters can be done by checking that it is not an outlier in diagrams based on quantities with
similar dependence with radius. 
Since both galaxies and dark matter are collisionless,
stellar mass is expected to be distributed as dark matter, 
as generally found in weak-lensing analyses of clusters at intermediate
redshift (e.g., Hoekstra et al. 1998, Kneib et al. 2003) and tested in detail for a few clusters 
(Andreon 2015). Therefore, most of
the stellar mass is at large cluster-centric radii, like the total mass.
We expect therefore a normal stellar fraction for CL2015.

Stellar masses have been derived from $r$ band luminosity of red-sequence galaxies within
$r_{500,cau}$ , as done for the clusters to which we compare 
CL2015 (Andreon 2010, 2012, to which we defer for details). In short, we
used $r<19$ mag photometry from the 14$^{th}$ Sloan
Digital Sky Survey (SDSS) data release (Abolfathi et al. 2018), we selected galaxies within
0.1 mag redward and 0.2 mag blueward of the $g-r$ color-magnitude relation, and we fit, without
binning, the luminosity function accounting for background, estimated in adjacent lines of sight outside
the cluster turnaround radius, 
assuming uniform priors for the parameters, a Schechter (1976) function for cluster galaxies, a power law for background
galaxies, and the likelihood expression in Andreon, Punzi \& Grado (2005). Since 
BCGs might not be drawn from the Schechter function because of their possibly
different formation history, we removed the BCG from the fitted sample and we added back its flux to the
integral of the luminosity function (as in Andreon 2010). Figure~\ref{LF} shows the counts in the cluster and
control field directions (top panel); the cluster contribution is given by the difference of these
two, shown in the middle panel. The BCG (leftmost point), IC1602, is about 2 mag brighter than other 
galaxies, which is the minimum difference needed to call the system a ``fossil" cluster (Jones et al. (2003)). 
In this respect, CL2015 is not unusual: among the 42 X-ray selected clusters studied in Andreon (2010),
six (about 15\%) have a similar magnitude difference with the other cluster galaxies.
In the much larger sample
of Lauer et al. (2014), IC1602 is $1\sigma$ brighter than the average BCG and 
its central velocity dispersion is in line with its 
central luminosity (i.e., it obeys the Faber \& Jackson 1976 relation when applied to central quantities).

The bottom panel of Fig.~\ref{LF} shows the contribution of each luminosity bin to the total flux, showing
that faint bins contribute little to the total flux, meaning that the data used are adequate for our purposes.
The total flux (the integral of L times the luminosity function) is then corrected by 15\% for missing light,  
converted to stellar
mass assuming the $M/L$ derived by Cappellari et al. (2006), and deprojected assuming
a Navarro et al. (1997) profile with a light concentration of three, as for the 
clusters to which we are comparing.  
We found a stellar mass of $\log M_{500,\ast}=12.50\pm0.08$ $M_\odot$.

Figure~\ref{stellar_fraction} compares CL2015 to other clusters: X-ray relaxed clusters 
(from Andreon 2012) and a 
random sampling of a X-ray flux-limited sample (from Andreon 2010), the latter binned
to emphasize the trend by reducing the
scatter around the mean relation. The fit to X-ray relaxed clusters (solid line) uses 
the same fitting code used in Andreon (2012). 
The stellar fraction of CL2015 within $\Delta=500$  (blue square) is consistent with
that of the other clusters.  This indicates that in spite its low concentration, CL2015 has a normal stellar
fraction.  The normal stellar
fraction is what we expected for a cluster with any
concentration because concentration is irrelevant for a comparison among 
quantities distributed as mass. The normal stellar fraction provides 
an independent test of our conclusion that concentration is driving the different shining of CL2015.

\begin{figure}
\centerline{\includegraphics[width=9truecm]{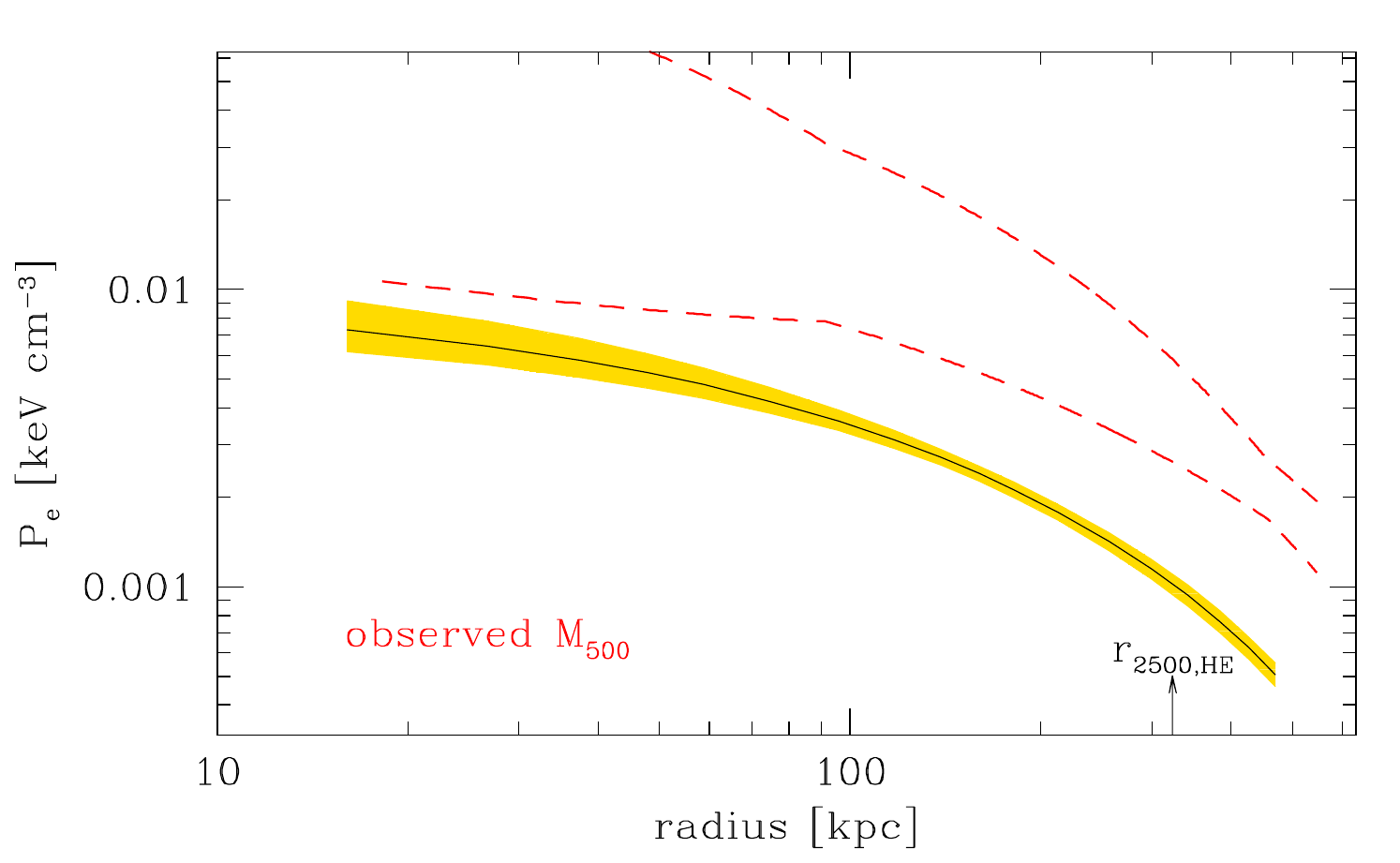}}
\centerline{\includegraphics[width=9truecm]{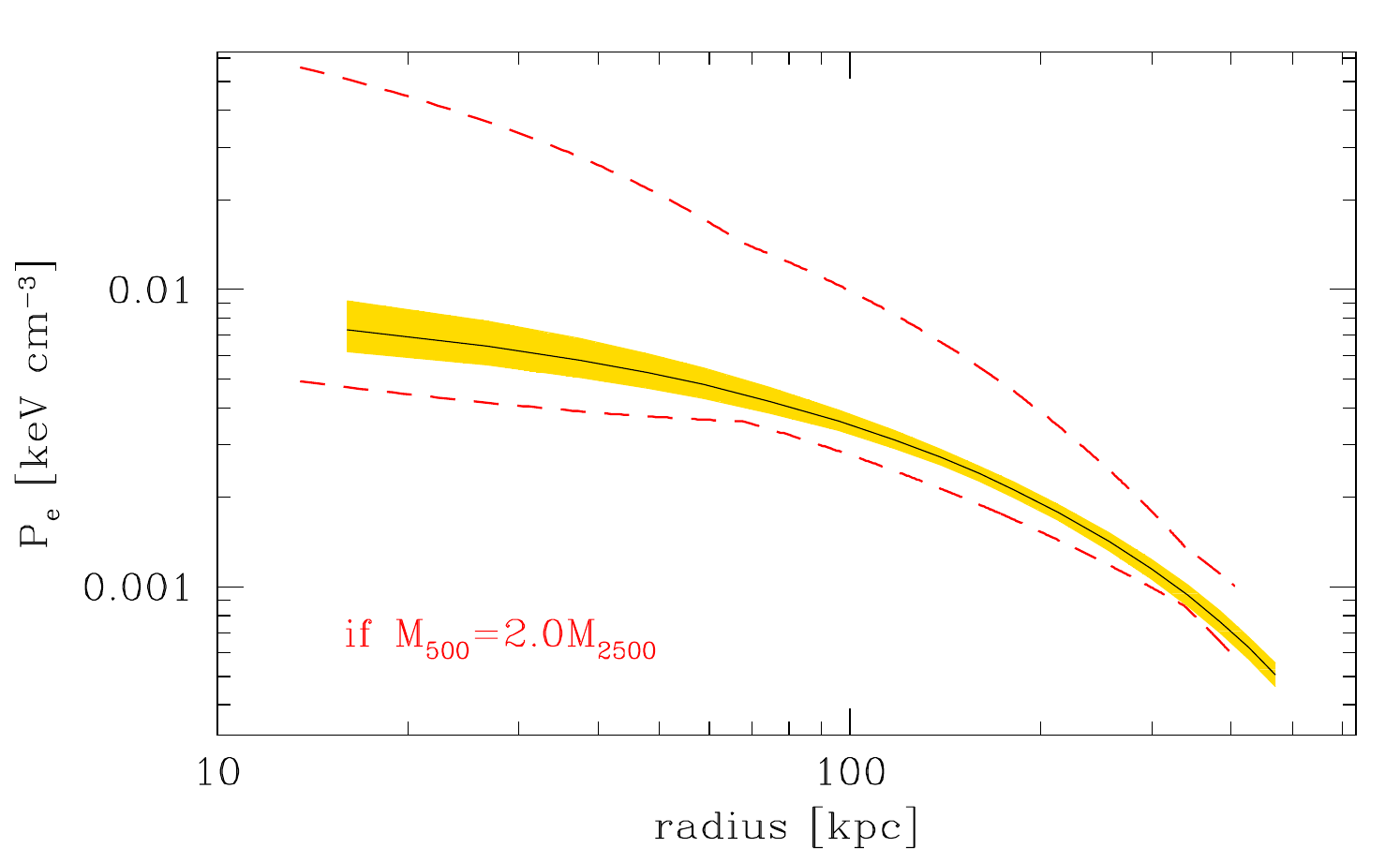}}
\caption{Pressure profile of CL2015 (line, the shading marks the 68\% uncertainty) vs. the $\pm2\sigma$ range of REXCESS clusters 
scaled at the CL2015 mass (top panel) and at the mass $M_{500}=2.0 M_{2500}$ (bottom panel) typical of clusters 
of normal concentration.}
\label{presscomp}
\end{figure}

\begin{figure}
\centerline{\includegraphics[width=9truecm]{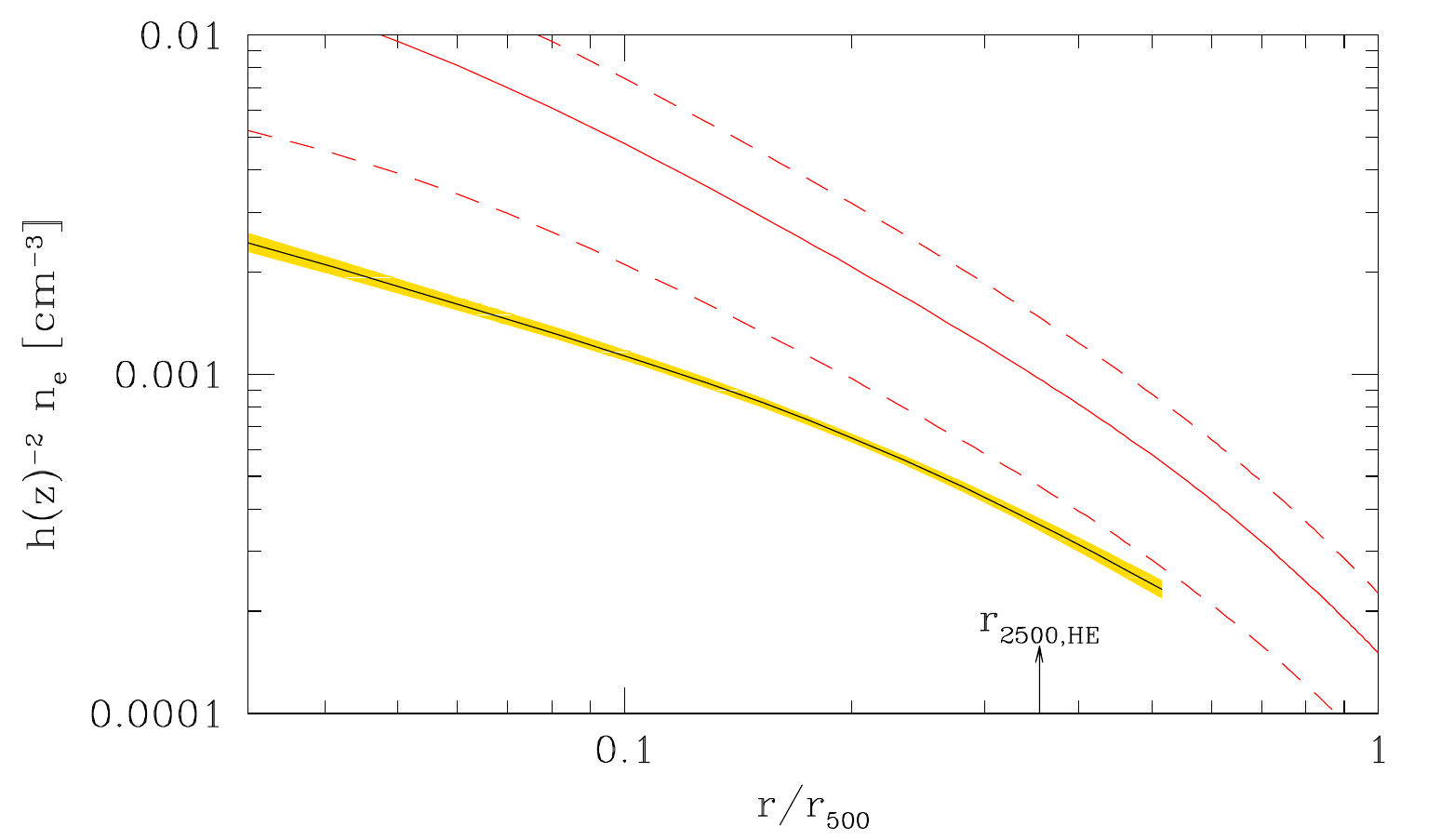}}
\centerline{\includegraphics[width=9truecm]{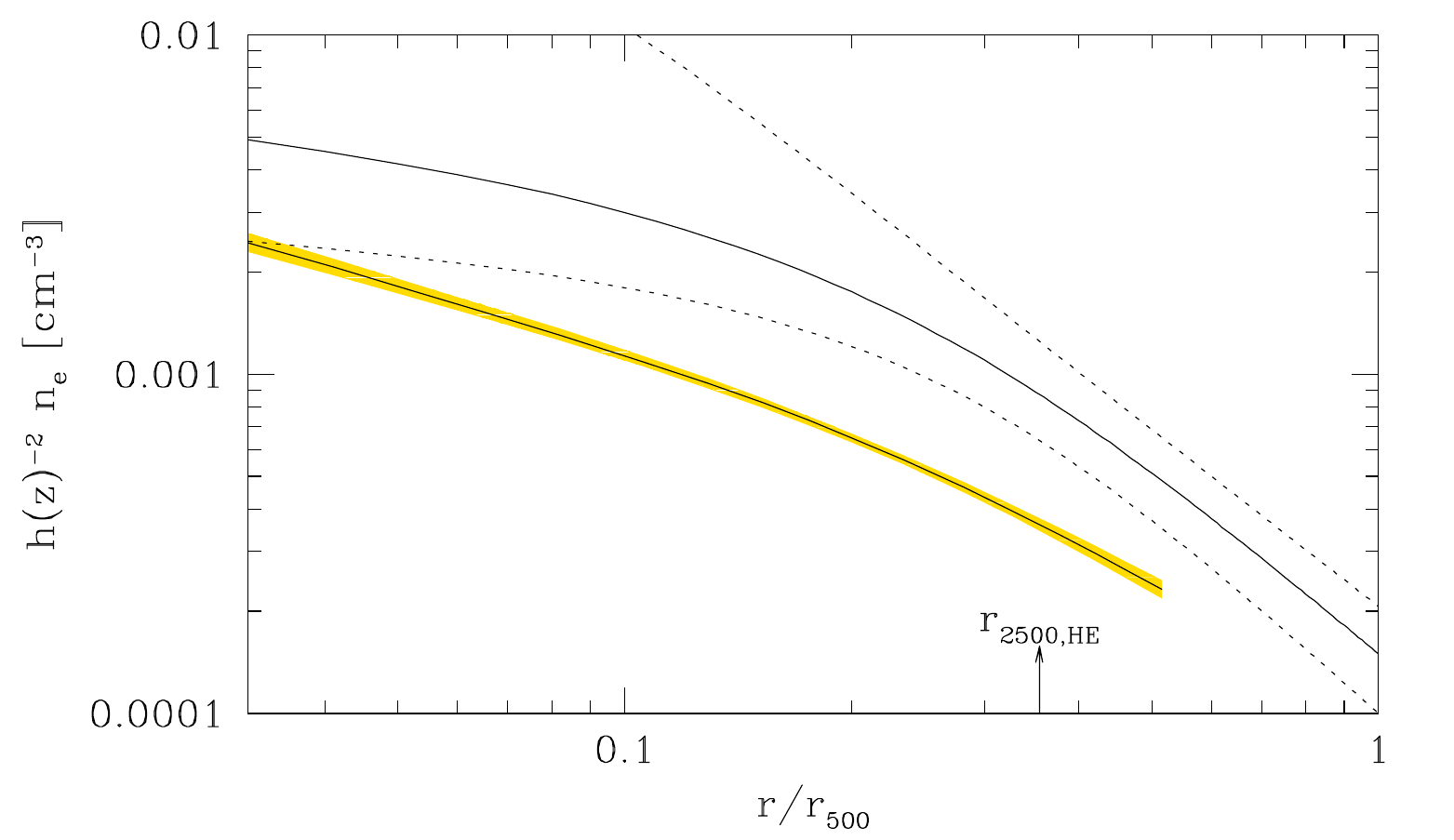}}
\centerline{\includegraphics[width=9truecm]{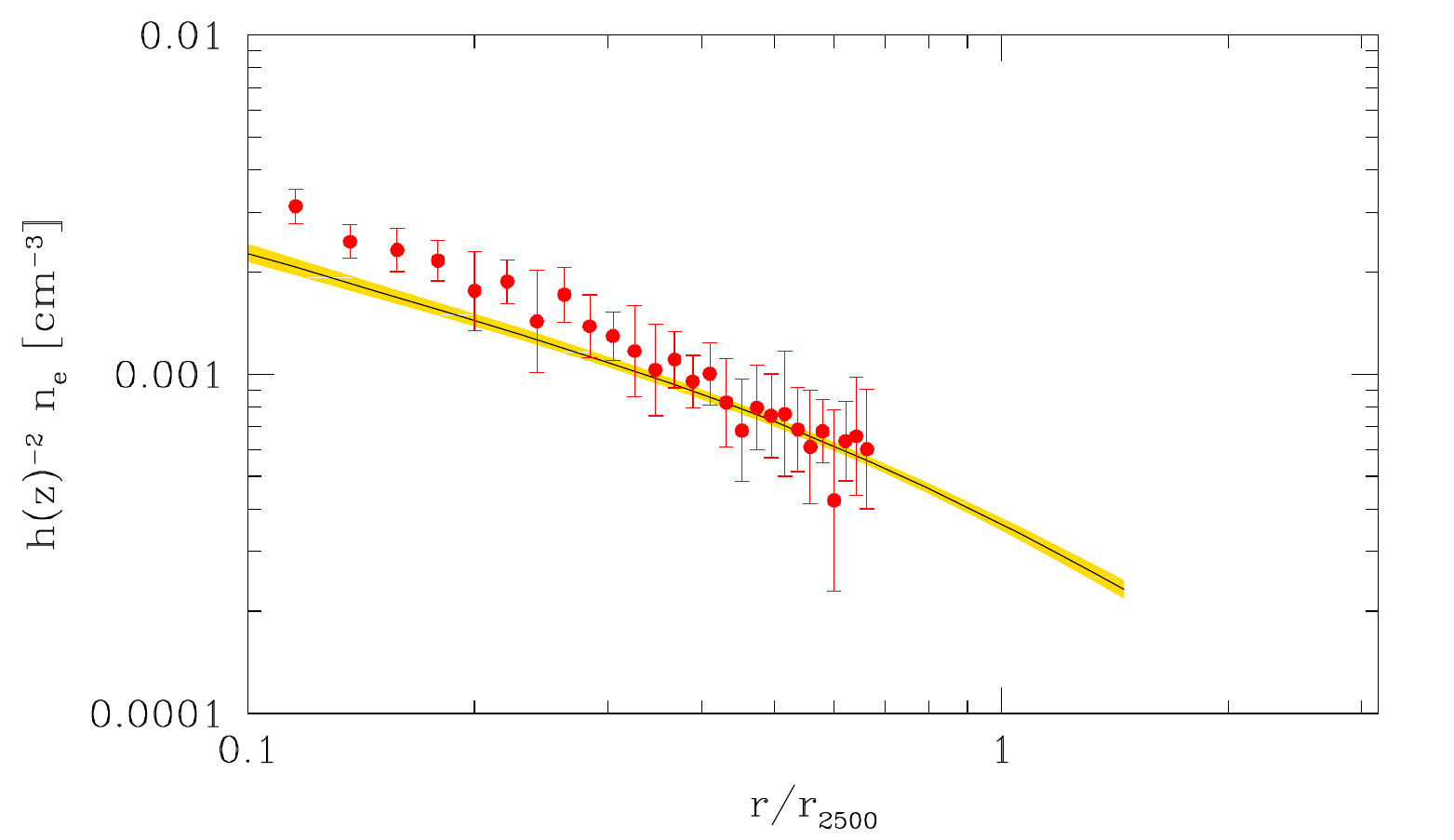}}
\caption{Electron pressure profile of CL2015 (line, the shading marks the 68\% uncertainty) vs. the $\pm2\sigma$ range of XCOP clusters 
(top panel), Planck-selected clusters (middle panel), and of Abell 744, a cluster with the same $M_{2500}$ (bottom panel). 
}
\label{necomp}
\end{figure}

\subsection{Pressure and density profile comparisons before and after factoring out the unusual concentration}

In this section we want to show that, once differences in concentration
are taken into account, CL2015 displays the same pressure profile as other clusters.
Figure~\ref{presscomp} compares the pressure profile of CL2015  (solid line with shading 
marking the 68\% uncertainty) with the 
universal pressure profile and its $\pm2\sigma$ range derived from REXCESS
clusters (Arnaud et al. 2010). In the top panel, we 
scaled the universal profile to the mass of CL2015 at overdensity $\Delta=500$.
CL2015's profile is systematically below the
$-2\sigma$ range, meaning it has a pressure systematically and significantly lower than any of the 
REXCESS clusters, hence breaking the universality of the pressure profile. 
However, the mass inside the radial range in which pressure is measured 
is different for CL2015 and REXCESS clusters because
$M_{500}=4.7 M_{2500}$ for CL2015 and $M_{500}=2.0 M_{2500}$ for the others. To
(approximatively) factor out the effect of concentration we therefore use 
$M_{500}=2.0 M_{2500}$ also for CL2015 (i.e., we effectively use $M_{2500}$ to rescale
all clusters). The bottom panel shows that CL2015's pressure profile is enclosed
in the $\pm2\sigma$ range of 
REXCESS clusters. The unusual pressure profile is largely driven by the unusual
concentration.

Figure~\ref{necomp} repeats much of the same exercise, but for the electron density profile:
the top panel compares the electron density profile of CL2015 to the universal density profile
and its $\pm2\sigma$ range derived from X-COP
clusters (Ghirardini et al. 2019). CL2015's profile is systematically below the
$-2\sigma$ range, that is, it is systematically and significantly lower than any of the 
X-COP clusters. 
We verified that the same holds true using a mass-matched sample of 19 clusters in 
REXCESS using electron density profiles in Croston et al. (2009). 
The middle panel shows that CL2015's electron density profile is even lower
than Planck-selected clusters (from Planck Collaboration 2011).

However, electron density is low because the compared clusters have a normal concentration while
CL2015 has a low concentration.
CL2015's electron density profile matches other cluster profiles within $r_{2500}$ if one
chooses a comparison cluster with the same mass within this overdensity, as illustrated by
the bottom panel where we compare CL2015 to Abell 744 (points from Cavagnolo et al. 2009; $r_{2500}$ from
Sun et al. 2009). This object has been selected as the object closest to CL2015 in 
entropy, temperature, and gas fraction among those in Fig.~\ref{CL2015vsSun}.
However, this cluster lacks $r_{500}$ estimates and therefore we ignore if it is another
cluster of low concentration, or a cluster of low $M_{500}$ mass.

\begin{figure}
\centerline{\includegraphics[width=9truecm]{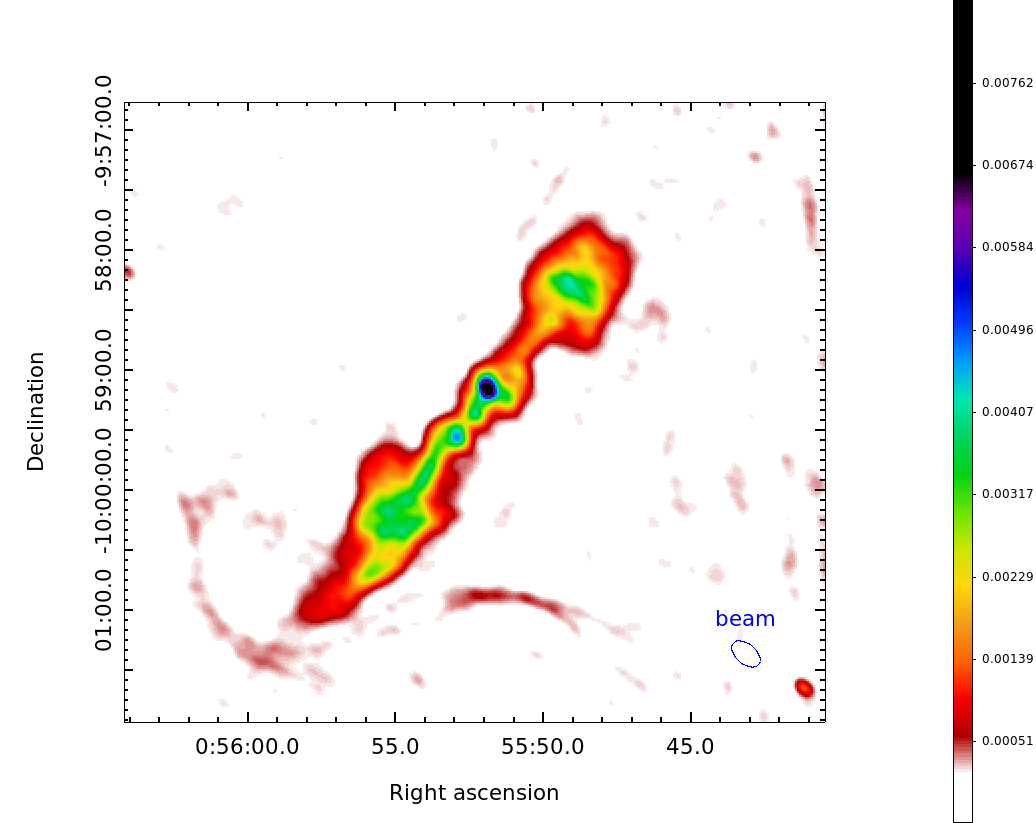}}
\caption{Giant Metrewave Radio Telescope  400 MHz image of the central part (about 0.1\% of the imaged 
area) of CL2015. The rms near the source is $\sim$ 100 $\mu$Jy/beam and the resolution of the 
image is 8.3 $\times$ 5.0 arcsec at 50 deg. position angle. The (optical) BCG position
is coincident with the brightest radio spot, at (ra,dec)=(0:55:51.9,-9:59:08). 
The color scale has Jy/beam units.}
\label{GMRTradioima}
\end{figure}

\subsection{Radio}

We observed CL2015 with the
upgraded Giant Metrewave Radio Telescope (uGMRT, Gupta et al. 2017)
in band-3 (250 - 500 MHz) for approximately three  net hours of
integration. The data were
recorded with 200 MHz bandwidth (300 $-$ 500 MHz)  and 4096 channels.
The data were analyzed
in a fully automated Common Astronomy Software Application-based pipeline (Ishwara-Chandra et al. 2019;
see also http://www.ncra.tifr.res.in/~ishwar/pipeline.html) using standard 
wideband analysis procedures, which include setting flux scale using the primary calibrator
3C48, bandpass, and gain calibration and transferring the flux scale to the science target
via the phase calibrator. After the gain calibration, the data were averaged with the post-averaged channel
width of $\sim$ 0.5 MHz before split, to keep the bandwidth smearing negligible.
The science target was imaged using the CASA task {\tt tclean} following wideband 
multi-frequency synthesis procedures. Four iterations of
phase-only self-calibration were performed, followed by four iterations of amplitude and phase self-calibration. During self-calibration,
flagging was performed on the residuals for a better result. The resolution and rms of the image are 8.3 $\times$ 5.0 arcsec at 50 deg. position angle
and $\sim$ 100 $\mu$Jy/beam near the central source.

\begin{figure}
\centerline{\includegraphics[width=7truecm]{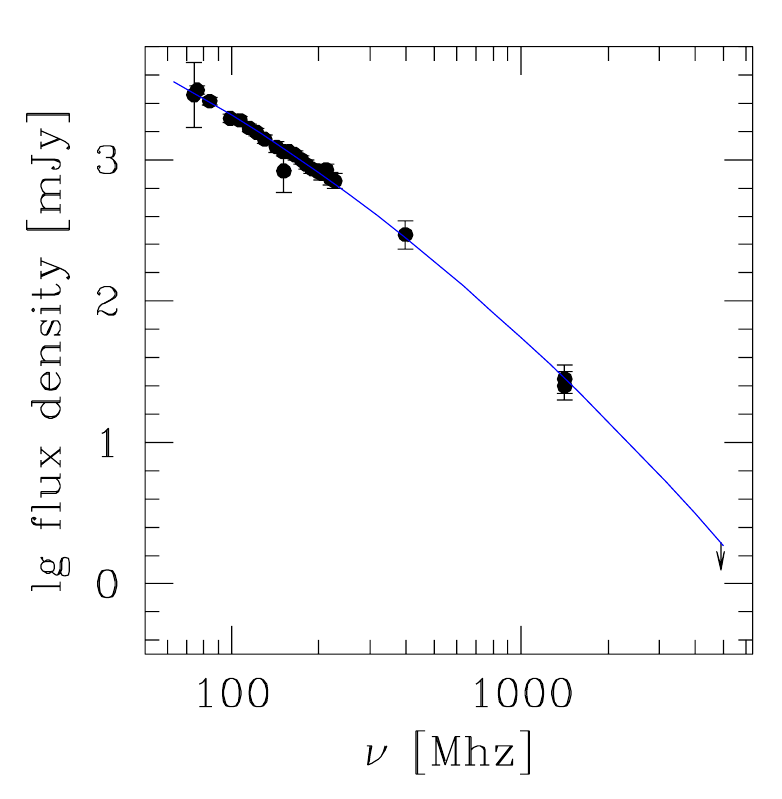}}
\caption{Radio spectrum of the BCG of CL2015, IC1602. The curve has the equation: $\log f =  -0.315934 \log^2\nu +0.00838119\log\nu + 4.56086.$}
\label{GMRTspectrum}
\end{figure}

The central radio source (Fig.~\ref{GMRTradioima}), spatially coincident with the BCG of CL2015, IC1602, 
is a double-lobed source with complex morphology.
The radio jets show several knots and
twists, which could be due to an inhomogenous medium.   
The linear extent of the source is about 120
kpc radius (about $r_{2500}/3$), the radio luminosity  at 400 MHz is 2 $\times 10^{24}$ WHz$^{-1}$, and 
the radio flux is $\sim$ 295 mJy. At 1.4 GHz IC1602 is faint, but still in the range
of other cluster radio sources, for example, those in Owen et al. (1997). 

When combined with GMRT observations at 153 MHz (Intema et al.
2017), Very Large Array Low-Frequency Sky Survey at 74 MHz (Cohen et al. 2007), Very Large Array at 1.4 GHz (Owen et al. 1997), 
Murchison Widefield Array (MWA) at several bands from 76
to 227 MHz (Hurley-Walker et al. 2017), NRAO VLA Sky Survey at 1.4 GHz (Condon et al. 1998), and the upper limit
at 4.8 GHz (Lin et al. 2009), the spectrum of the whole source
shows a curvature with a clear steepening towards high frequencies (Fig.~\ref{GMRTspectrum}).
The spectral index at the low frequency part (74 MHz till 400 MHz) is $\sim$ 1.3, whereas it is $\sim$ 2 in the frequency 
range 400 MHz $-$ 4.86 GHz.  Such a curved spectrum with steep spectral index is typical 
of source relics  (e.g., Brienza et al 2016). We suspect the radio source has undergone episodic
jet activity and the relic plasma from the previous jet activity is
primarily radiating at low radio frequencies causing the excess
flux. The presence of diffuse emission beyond the ``hotspots" on either
side could be the relic emission from the active galactic nucleus which is responsible
for steep spectra.

Although an AGN is present at the cluster center and we see radio lobes, its effect on the X-ray gas  is not
discernable in our X-ray data: we see no cavities, no rims surrounding them, no sharp break in the surface
brightness profile, no azimuthal asymmetry, all of which are seen, for example, in Abell 2390 (Vikhlinin et al. 2006), however
using higher resolution Chandra data.  The presence of an AGN at the center of the cluster might lower the
gas fraction in the cluster center by displacing the gas at larger radii (but it should produce X-ray cavities that we
do not see). It might also bias  our estimate of the gas fraction to the high end if it provides non-thermal pressure
support to the gas (not accounted for in our hydrostatic analysis). Given that gas fraction is normal at
$\Delta=2500$ (Sect.~3.4), and also consistent with Vikhlinin et al. (2006) clusters of the same
temperature, deviations from hydrostatic equilibrium should be small unless there is fine tuning between
the amount of gas displaced and the mass overestimate.

An interesting feature is the "S"-shaped narrow but long shock-like feature at the end of southeastern lobe
(see Fig.~\ref{GMRTradioima}).
This is clearly  disconnected from the rest of the radio source. The detailed investigation of this arc
requires much deeper multiwavelength observations and is beyond the scope of this paper.

\begin{figure}
\centerline{\includegraphics[width=9truecm]{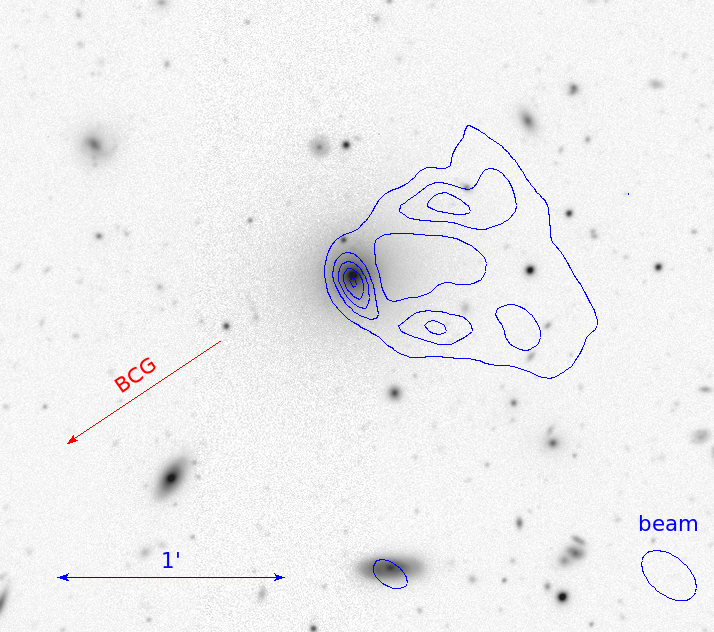}}
\caption{DECam Legacy survey (Dey et al. 2018) $r$-band image with GMRT 400 MHz contours of
the head tail source. The galaxy has J2000 coordinates (ra,dec)=(0:55:32.5,-9:56:27). North is up and east is 
to the left. Contours are in steps of 1 mJy/beam 
starting from 0.3 mJy/beam.
}
\label{GMRTinfallsource}
\end{figure}

Figure~\ref{GMRTinfallsource} shows
a low luminosity head tail source 5.5' arcmin northwest of the
cluster center, centered on a spectroscopically confirmed cluster member (at 0:55:32.5,-9:56:27).
The flux density of the source at 400 MHz
is 53 mJy. The source is also detected in NVSS (Condon et al. 1998) with a flux density of
23.9 mJy. The source is barely detected in MWA (Hurley-Walker et al. 2017) and not detected in GMRT 
Sky Survey (Intema et al.
2017) and this gives the spectral index of $\sim$ 0.7.
The large field of view of the GMRT image allows us to easily
spot at least five more clusters in the field of view of $\sim$ 2 deg$^2$ 
centered on CL2015, which will be reported elsewhere.

\section{Discussion}

Cluster mass can be estimated in several ways, for example, from the escape velocity,
which does not make assumptions on cluster status
(caustics, Diaferio 1999; Serra et al. 2011), from velocity dispersions
assuming dynamical equilibrium, or from X-ray data assuming hydrostatic equilibrium. 
If the cluster is out
of equilibrium, these estimates tend to disagree with each other. In particular, in a system
that has experienced a recent
merger, mass estimates from velocity dispersion and caustic should
differ each other, except if the merger occurs close to the plane of the sky. 
Mergers close to the plane of the sky
are those easier to recognize from the unrelaxed
X-ray cluster morphology (e.g., Takizawa et al. 2010) and from
the positional offset between the (collisional) gas and the (collisionless) galaxies. 
Independently from the merger direction, in out-of-equilibrium clusters
the X-ray mass based on hydrostatic equilibrium tends to underestimate the true
mass because of the neglected pressure support (e.g., Rasia et al. 2004, Takizawa et al. 2010).
Therefore, X-ray morphology and
different estimates of mass allow us to
recognize the presence of a recent merger.  Both the regular X-ray morphology of CL2015 and the
coincidence of the X-ray center with the BCG location  exclude a merger close to 
the plane of the sky. The agreement between masses derived from caustic and
velocity dispersion (most sensitive to
a merger along the line of sight) excludes a merger with a component along the line of sight.
Finally, the agreement between hydrostatic and dynamical masses excludes a strong non-thermal
support, including a merger. A merger along the line of sight
is also the easiest to recognize in redshift-space diagrams 
and none is seen (also shown in Fig. 9 of Abdullah et al. 2018). Therefore,
a recent merger can be excluded, and cannot be invoked to explain
the low X-ray emission and low concentration
of CL2015.
While not viable for CL2015, mergers could provide an explanation for
other XUCS clusters with low X-ray emission, for example for CL3000, which shows
an offset between the gas and the galaxies, a bit like the Bullet cluster (Andreon et al. 2016).

Neither the presence of a bright BCG, nor the cluster being classified as fossil, explain
the low X-ray emission and low concentration
of CL2015, because other clusters in XUCS are X-ray faint for their mass but do not
have a similar luminosity difference between the first and second members, and because the
BCG has a luminosity typical of other BCGs.

CL2015 is perhaps the first known cluster with a mass concentration as low as $c_{500}=1.5$. 
It should not be seen, however, as an anomaly: the rarity of such objects in X-ray selected
observational samples is not an indication of their rarity in
the real Universe because their low central mass induces a very low X-ray luminosity that 
makes them hard to include in observational X-ray selected samples. 
On the other hand, CL2015 is just 
 $1.5\sigma$ away from the mean luminosity for its mass in an X-ray unbiased
sample, meaning it is not unusual at all. 
It is interesting to note that the shear-selected
sample of Miyazaki et al. (2018) has an average concentration of $c_{500}=2.4\pm0.5$, lower
than the common $c_{500}=4.2$ value obtained for X-ray selected clusters, hence supporting
the idea that low concentration clusters are more common in samples not affected by the X-ray bias.
One might be tempted to define  CL2015 as "underluminous" when compared to X-ray clusters.
However, this label assumes that clusters such  as CL2015 are the exception
in the Universe,  while  X-ray selected clusters represent the norm.
When compared to an X-ray unbiased sample, CL2015 is just $1.5\sigma$ away from the mean luminosity
for its mass, therefore cannot be labeled ``underluminous''.   Recent works (see Introduction) 
now agree that clusters that enter more easily
in X-ray-selected samples are "too bright"  (e.g., virtually all clusters
in the complete,  X-ray selected sample of Giles et al. 2017 are brighter than the average
for their mass).

Among the new cluster population
that has been discovered in recent years, 
CL2015 has a unique dataset.
Pearson et al. (2017) presents several optically-selected 
groups that are X-ray faint for their mass, 
where however these mass estimates comes from galaxy counts (similar to Andreon \& Moretti 2011).
Better estimates of mass are required to confirm these clusters as truly faint for their mass. 
Similarly, Planck collaboration (2016) and Rossetti et al. (2017) found clusters
possibly underluminous for their integrated pressure parameter, but they need
masses (and also deeper X-ray data in the case of Planck collaboration 2016)
to confirm them as faint for their mass, or as having a low mass concentration. 
Xu et al. (2018) discovered groups with 
a low surface brightness X-ray emission 
(more precisely, more extended than a cluster model scaled down to have the observed 
group luminosity). However, lacking precise mass estimates
they cannot investigate the reason for the low surface brightness, whether this
is due to a low total mass concentration or to a reduced gas fraction in the center.
Furthermore, because of the shallowness of the X-ray data, the detection
of clusters of low surface brightness like CL2015 is precluded to them, and indeed
CL2015 is not in their sample.
As a consequence, their detected objects are expected to sample a narrower range
in surface brightness than the XUCS sample.

\section{Conclusions}

In recent years, the known variety of galaxy clusters has been constantly increasing, as
testified by the increasing scatter displayed by the scaling relations, by a larger
spread in gas fraction, and by the discovery of clusters with low electron density profiles.
In this context, we obtained a 58 ks 
Swift XRT observation, a GMRT band-3 10.8 ks  exposure, and we collected archival data from 
Einstein, Planck, and SDSS on CL2015,
an intermediate mass cluster ($\log M_{500}/M_\odot=14.39\pm0.09$) 
in the nearby Universe. The cluster was selected 
because it has a low X-ray luminosity ($L_{X,500,ce}=42.79\pm0.06$
in the [0.5,2] keV band) for its mass $M_{500}$, 
about $\sim12\sigma$ below the mean relation for an X-ray selected sample,
yet only $\sim1.5\sigma$ below the mean of an X-ray unbiased sample.

CL2015 is not an outlier because of faulty X-ray luminosities,  
because values derived by different X-ray telescopes and teams
agree with each other. Faulty masses are also not the cause, because the hydrostatic mass we
derive from XRT agrees with three dynamical estimates derived by two independent 
teams, and because CL2015 is also an outlier in a scaling that replaces mass with
an integrated pressure parameter. A recent merger is not the cause, because 
such an event is ruled out by X-ray and dynamical data. 

Universal profiles are deeply rooted in our understanding of
clusters. For example, the Planck cluster detection algorithm assumes a
universal pressure profile for all clusters, and pressure (the integral
of the pressure profile) is perhaps the current most appealing mass proxy. 
The pressure and the electron density profiles of CL2015 are systematically outside
the $\pm 2\sigma$ range of the universal profiles, with the electron density profile being
even lower than that of Planck-selected clusters.  CL2015 turns out also to be an outlier
in the X-ray luminosity versus SZ strength, but a normal object in terms of
stellar mass fraction. 

CL2015's hydrostatic mass profile, by itself or when it is considered together
with dynamical masses, shows that
the cluster has a remarkably low concentration and different sparsity
compared to clusters in X-ray selected samples.
CL2015 has a $c_{500}=1.5\pm0.4$ versus 
a typical $c_{500}=4.2,$ and $M_{500}/M_{2500}=4.7$ versus a typical value of $2.0$.
The unusual concentration and (inverse) sparsity makes CL2015 an outlier in scaling relations involving integrated
quantities with different radial dependencies,  such as $L_{X,500,ce}-M_{500}$
and $L_X-Y_{SZ}$, even when they are measured inside the same radial range. 
This is confirmed by the evidence that CL2015 is not an outlier 
in relations involving quantities with similar dependencies with radius, like 
stellar mass versus total mass. Furthermore, when concentration differences are accounted for,  
the properties of CL2015 become consistent with comparison samples, for example, enclosed gas fraction, 
X-ray temperature, and entropy at 
$R_{2500}$, and the pressure profile falls in the range of the ``universal''
profile derived for X-ray selected clusters. 

The different sensitivity of various observables to radius 
is promising for the collection of larger samples of low concentration clusters, for example 
looking for outliers combining quantities mostly sensitive to small radii, such as
$L_X$, to quantities which instead continue to grow with increasing radius (or decreasing
$\Delta$), such as total mass, SZ strength, or stellar mass.

CL2015 is perhaps the first known cluster with such a low concentration and high
quality data. It should
not be seen, however, as an anomaly: the rarity of such objects in
observational samples is not an indication of their infrequency in 
the real Universe because their low X-ray luminosity makes it difficult for them 
to be included in observational samples. As we discuss, while CL2015 is an 
outlier 
relative to X-ray selected samples, its low luminosity is shared by other clusters
of similar mass in the X-ray unbiased sample of Andreon et al. (2016), and 
a few other  X-ray faint clusters relative to their SZ strength  in  
Planck Collaboration (2016: one such cluster is indeed CL2015).

\begin{acknowledgements}
We warmly thank Ming Sun for useful discussions, Jeremy Sanders
for  giving us support in the use of MBPROJ2, and Gary Mamon 
for help with the $r_{200,DM}-r_{200}$ conversion. 
We thank Ming Sun, Iacopo Bartalucci, and David Barnes for giving us tabulated radial profiles
and individual values plotted in figures of their papers. We thank Fabio 
Gastaldello, Ivan De Martino, Mariachiara Rossetti, and Franco Vazza for discussions
on CL2015.
We acknowledge the financial contribution from the agreement ASI-INAF n.2017-14-H.0 and
PRIN MIUR 2015 Cosmology and Fundamental Physics: Illuminating the Dark Universe  
with Euclid.
We thank the staff of the GMRT that made these observations possible. 
GMRT is run by the National Centre for Radio Astrophysics of the Tata Institute of Fundamental Research.
\end{acknowledgements}

{}

\appendix

\section{Mass from galaxy dynamics}

\begin{figure}
\centerline{\includegraphics[width=6truecm]{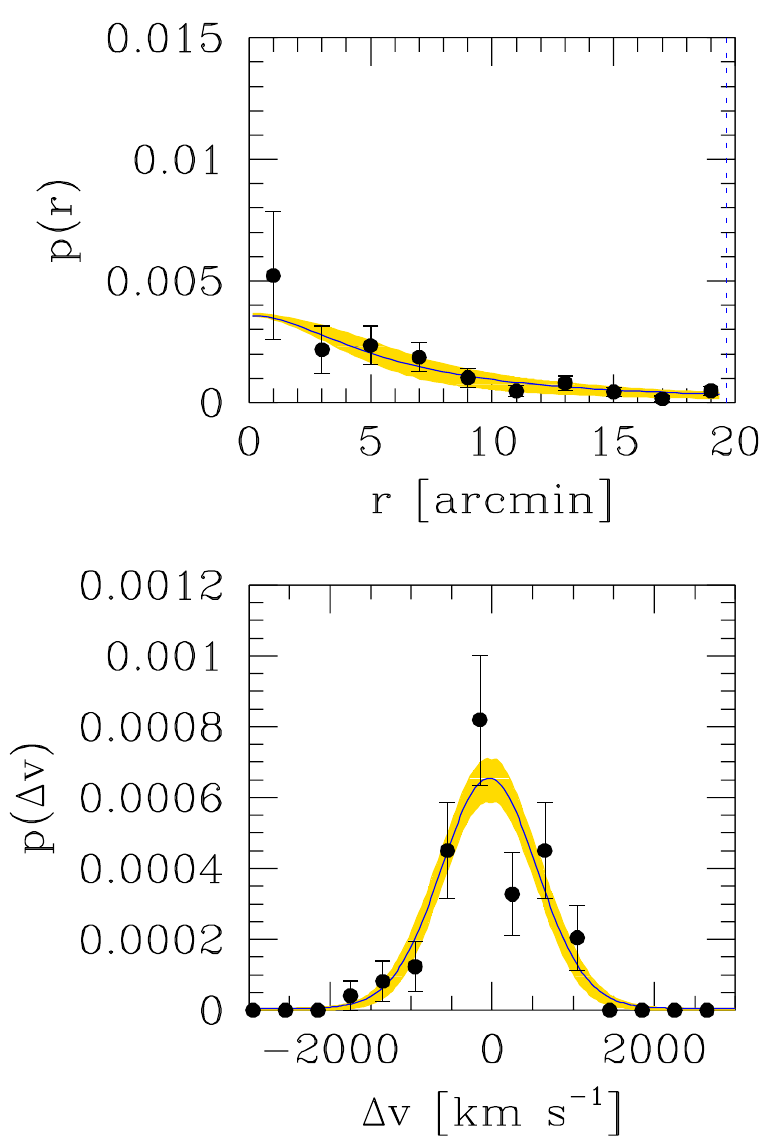}}
\caption{Spatial and velocity distributions. 
{\it Top panel:} Radial profile of the galaxies with spectroscopic redshift within 3000 km/s of CL2015's
redshift.
{\it Bottom panel:} Velocity distribution of the same galaxies. In both panels, the solid line
is the mean model fitted to the individual galaxy data in the $r-v$ plane, 
while the shading indicates the 68\% uncertainty
(highest posterior density interval). Points and approximated
error bars are derived by ignoring the velocity (top panel) or radial (bottom panel) information, 
binning galaxies in velocity bins, and adopting approximated Poisson   
errors, as is commonly found in the literature.} 
\label{CL2015dyn}
\end{figure}

CL2015's dynamical masses have been derived in three different ways by two
different teams, Abdullah et al.
(2018) and Andreon et al. (2016), who, however, do not show figures
for CL2015.
These are shown here for completeness.

In Andreon et al. (2016) we computed the cluster velocity dispersion
with a full Bayesian modeling of
the data in the radius-velocity space, allowing a Gaussian velocity and
a King radial profile for
member galaxies and a uniform distribution in velocity and radial distance
for foreground and background galaxies (see Andreon et al.
2016 for details).  
There are 61 spectroscopic members in CL2015. The marginal distributions are shown in
Fig.~\ref{CL2015dyn}. The
cluster members are adequately described by the model, with $\sigma_v=
595 \pm 56$ km s$^{-1}$, from which
$M_{200}$ is derived using the calibration in Evrard et al. (2008), and
then $M_{500}$ assuming
a Navarro et al. (1997)'s profile with concentration 5. We found $\log
M_{500,\sigma_v}=14.27\pm0.12$ (Andreon
et al. 2016).

Caustic masses within $r_{200}$, $M_{200}$, have been derived following
Diaferio \& Geller
(1997), Diaferio (1999), and Serra et al. (2011), then converted into
$r_{500}$ and $M_{500}$
assuming a Navarro, Frenk, \& White (1997) profile with concentration 5.
We found $\log M_{500,cau}=14.36\pm0.09$ (Andreon
et al. 2016).

The GalWeigh technic (Abdullah et al. 2018) uses much of the same
information used in the caustic mass
determination (velocity and cluster-centric distance) for identifying
cluster members,
from which cluster masses can be derived with different sets of assumptions.
Assuming a Navarro, Frenk, \& White (1997) profile and using
spectroscopic data that overlap with those
used by us, the authors found
$\log M_{500,dyn}=14.27$ (no error quoted) for CL2015. The caustic diagram
of CL2015 (named A117 in their paper) is shown in their Fig.~9.

\section{Test of our reduction and analysis pipeline}

\begin{figure}
\centerline{\includegraphics[width=9truecm]{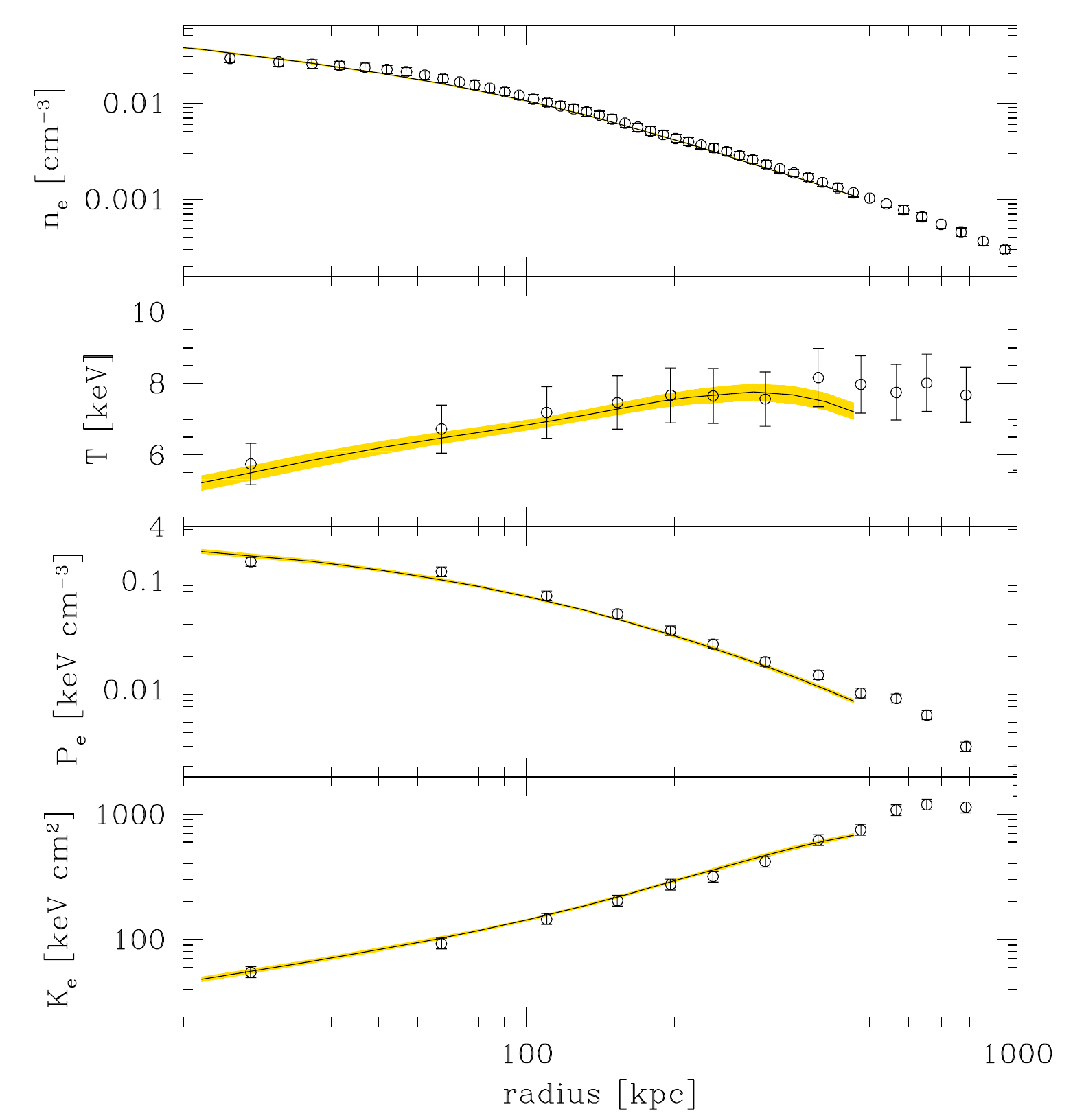}}
\caption{Comparison of XMM and XRT Abell 2029 thermodynamic profiles. Points indicate values derived
from XMM (Ghirardini et al. 2019) while error bars indicate $\pm10\%$ (statistical errors are negligible), 
whereas our XRT mean model is indicated
with a solid line and shading (marking 68\% intervals).}
\label{A2029phys}
\end{figure}

To test our ability to derive thermodynamic profiles versus 
state-of-the-art analyses,
Fig.~\ref{A2029phys} compares the thermodynamic profiles of Abell 2029
derived by us using
58 ks of Swift data with those derived
using $\sim 170$ ks of XMM-Newton data by Ghirardini et al. (2019).
The angular size of Abell 2029 exceeds the XRT field of view and
its high brightness largely dominates over the background even at the
boundary. 
However, since the computation of three-dimensional profiles at the radius $r$
needs to account for the emission at larger radii $r'>r$, and those
are not available when outside the field of view, radii close to the
boundaries are biased and we restricted the analysis to $r<500$ kpc for Swift.
In the XMM-Newton dataset for A2029 there are additional pointings covering the emission
at large radii, and therefore the profiles are reliable out to larger
radii. CL2015 is less affected by this problem because it has a smaller angular
(sky) dimension than Abell 2029. 

Broadly speaking, the three-dimensional thermodynamic profiles are
derived using similar methods
by us and Ghirardini et al. (2019) using both spatial and spectral data, but
the details are
different.
Our treatment of the background is simpler because a more complex
treatment
is unnecessary for the stable and low XRT background (except at special locations,
such as at low Galactic latitudes).
Ghirardini et al. (2019) assume no
temperature gradients in the derivation of the electron density profile and
during the Abel inversion
the number of photons in the annuli is computed from the
median-averaged profile, while it is given by the mean-average profile
(times the shell area) because regions
of higher emissivity are not removed in their spectral analysis.
Especially at the center, where the emitting volume is small
and embedded in bright cluster emission, the XCOP choice underestimates
the contamination
of the outer shells.

In spite of these differences,
profiles from different telescopes, analyses, and authors are within
10\% (Fig.~\ref{A2029phys}).
This
indicates that (telescope, analysis, and authors) systematics are well
below the 10\% level
and that our derivation of thermodynamic profiles
is sophisticated enough compared to other state-of-the-art analyses.
Our physical fit to Swift data gives
$\log M(<\rm{0.5 Mpc})/M_{\odot}= 14.36\pm0.02$ versus $14.44\pm0.005$
quoted in Ettori et al. (2018) from their XMM analysis,
where errors are
purely formal because deviations from the hydrostatic equilibrium
hypothesis alone are
at least on the order of $0.05-0.1$ dex, as quoted in Ettori et al. (2019).


\begin{thebibliography}{}

\bibitem[Abdullah et al.(2018)]{2018ApJ...861...22A} 
Abdullah, M.~H., Wilson, G., \& Klypin, A.\ 2018, \apj, 861, 22 

\bibitem[Abell(1958)]{1958ApJS....3..211A} Abell, G.~O.\ 1958, \apjs, 3, 211

\bibitem[Abolfathi et al.(2018)]{2018ApJS..235...42A} 
Abolfathi, B., Aguado, D.~S., Aguilar, G., et al.\ 2018, \apjs, 235, 42 

\bibitem[Anders \& Grevesse(1989)]{1989GeCoA..53..197A} 
Anders, E., \& Grevesse, N.\ 1989, \gca, 53, 197 

\bibitem[Andreon(2010)]{2010MNRAS.407..263A} 
Andreon, S.\ 2010, \mnras, 407, 263 

\bibitem[Andreon(2012)]{2012A&A...548A..83A} 
Andreon, S.\ 2012, \aap, 548, A83 

\bibitem[Andreon(2015)]{2015A&A...575A.108A} 
Andreon, S.\ 2015, \aap, 575, A108 

\bibitem[Andreon \& Moretti(2011)]{2011A&A...536A..37A} 
Andreon, S., \& Moretti, A.\ 2011, \aap, 536, A37 

\bibitem[Andreon \& Hurn(2013)]{2013SADM....6...15A} 
Andreon, S., \& Hurn, M.\ 2013, Statistical Analysis and Data Mining 9, 15

\bibitem[Andreon et al.(2005)]{2005MNRAS.360..727A} 
Andreon, S., Punzi, G., \& Grado, A.\ 2005, \mnras, 360, 727 

\bibitem[Andreon et al.(2011)]{2011MNRAS.412.2391A} 
Andreon, S., Trinchieri, G., \& Pizzolato, F.\ 2011, \mnras, 412, 2391 

\bibitem[Andreon et al.(2009)]{2009A&A...507..147A} 
Andreon, S., Maughan, B., Trinchieri, G., \& Kurk, J.\ 2009, \aap, 507, 147 

\bibitem[Andreon et al.(2016)]{2016A&A...585A.147A} 
Andreon, S., Serra, A.~L., Moretti, A., \& Trinchieri, G.\ 2016, \aap, 585, A147 


\bibitem[Andreon et al.(2017)]{2017A&A...606A..24A} % variegate gas content
Andreon, S., Wang, J., Trinchieri, G., Moretti, A., \& Serra, A.~L.\ 2017, \aap, 606, A24 

\bibitem[Arnaud et al.(2005)]{2005A&A...441..893A} 
Arnaud, M., Pointecouteau, E., \& Pratt, G.~W.\ 2005, \aap, 441, 893 

\bibitem[Arnaud et al.(2010)]{2010A&A...517A..92A} 
Arnaud, M., Pratt, G.~W., Piffaretti, R., et al.\ 2010, \aap, 517, A92 

\bibitem[B{\"o}hringer et al.(2004)]{2004A&A...425..367B} 
B{\"o}hringer, H., Schuecker, P., Guzzo, L., et al.\ 2004, \aap, 425, 367 

\bibitem[Brienza et al.(2016)]{2016A&A...585A..29B} 
Brienza, M., Godfrey, L., Morganti, R., et al.\ 2016, \aap, 585, A29 

\bibitem[Cappellari et al.(2006)]{2006MNRAS.366.1126C} 
Cappellari, M., et  al.\ 2006, MNRAS, 366, 1126 

\bibitem[Cavagnolo et al.(2009)]{2009ApJS..182...12C} 
Cavagnolo, K.~W., Donahue, M., Voit, G.~M., \& Sun, M.\ 2009, \apjs, 182, 12 

\bibitem[Cohen et al.(2007)]{2007AJ....134.1245C} 
Cohen, A.~S., Lane, W.~M., Cotton, W.~D., et al.\ 2007, \aj, 134, 1245 

\bibitem[Condon et al.(1998)]{1998AJ....115.1693C} 
Condon, J.~J., Cotton, W.~D., Greisen, E.~W., et al.\ 1998, \aj, 115, 1693 

\bibitem[Corasaniti et al.(2018)]{2018ApJ...862...40C} 
Corasaniti, P.~S., Ettori, S., Rasera, Y., et al.\ 2018, \apj, 862, 40 

\bibitem[Croston et al.(2009)]{2009MNRAS.395.1999C} 
Croston, J.~H., Kraft, R.~P., Hardcastle, M.~J., et al.\ 2009, \mnras, 395, 1999 

\bibitem[Dey et al.(2018)]{2018arXiv180408657D} 
Dey, A., Schlegel, D.~J., Lang, D., et al.\ 2018, ApJ, submitted (arXiv:1804.08657) 

\bibitem[Diaferio(1999)]{1999MNRAS.309..610D} 
Diaferio, A.\ 1999, \mnras, 309, 610 

\bibitem[Diaferio \& Geller(1997)]{1997ApJ...481..633D} 
Diaferio, A., \& Geller, M.~J.\ 1997, \apj, 481, 633 

\bibitem[Dolag et al.(2004)]{2004A&A...416..853D} 
Dolag, K., Bartelmann, M., Perrotta, F., et al.\ 2004, \aap, 416, 853 

\bibitem[Dressler \& Shectman(1988)]{1988AJ.....95..985D} 
Dressler, A., \& Shectman, S.~A.\ 1988, \aj, 95, 985 

\bibitem[Dwarakanath et al.(2018)]{2018MNRAS.477..957D} 
Dwarakanath, K.~S., Parekh, V., Kale, R., \& George, L.~T.\ 2018, \mnras, 477, 957 

\bibitem[Ettori et al.(2019)]{2019A&A...621A..39E} 
Ettori, S., Ghirardini, V., Eckert, D., et al.\ 2019, \aap, 621, A39

\bibitem[Evrard et al.(2008)]{2008ApJ...672..122E} 
Evrard, A.~E., Bialek, J., Busha, M., et al.\ 2008, \apj, 672, 122 

\bibitem[Faber \& Jackson(1976)]{1976ApJ...204..668F} 
Faber, S.~M., \& Jackson, R.~E.\ 1976, \apj, 204, 668 

\bibitem[Fanaroff \& Riley(1974)]{1974MNRAS.167P..31F} 
Fanaroff, B.~L., \& Riley, J.~M.\ 1974, \mnras, 167, 31P 

\bibitem[Ge et al.(2019)]{2019MNRAS.484.1946G} 
Ge, C., Sun, M., Rozo, E., et al.\ 2019, \mnras, 484, 1946.

\bibitem[Ghirardini et al.(2019)]{2019A&A...621A..41G} 
Ghirardini, V., Eckert, D., Ettori, S., et al.\ 2019, \aap, 621, A41
 
\bibitem[Giles et al.(2015)]{2015MNRAS.447.3044G} 
Giles, P.~A., Maughan, B.~J., Hamana, T., et al.\ 2015, \mnras, 447, 3044.

\bibitem[Giles et al.(2016)]{2016A&A...592A...3G} 
Giles, P.~A., Maughan, B.~J., Pacaud, F., et al.\ 2016, \aap, 592, A3 

\bibitem[Giles et al.(2017)]{2017MNRAS.465..858G} 
Giles, P.~A., Maughan, B.~J., Dahle, H., et al.\ 2017, \mnras, 465, 858 

\bibitem[Hoekstra et al.(1998)]{1998ApJ...504..636H} 
Hoekstra, H., Franx, M., Kuijken, K., \& Squires, G.\ 1998, \apj, 504, 636 

\bibitem[Hurley-Walker et al.(2017)]{2017MNRAS.464.1146H} 
Hurley-Walker, N., Callingham, J.~R., Hancock, P.~J., et al.\ 2017, \mnras, 464, 1146 

\bibitem[Intema et al.(2017)]{2017A&A...598A..78I} 
Intema, H.~T., Jagannathan, P., Mooley, K.~P., \& Frail, D.~A.\ 2017, \aap, 598, A78 

\bibitem[Jones \& Forman(1999)]{1999ApJ...511...65J} 
Jones, C., \& Forman, W.\ 1999, \apj, 511, 65 

\bibitem[Jones et al.(2003)]{2003MNRAS.343..627J} 
Jones, L.~R., Ponman, T.~J., Horton, A., et al.\ 2003, \mnras, 343, 627 

\bibitem[Kalberla et al.(2005)]{2005A&A...440..775K} 
Kalberla, P.~M.~W., Burton, W.~B., Hartmann, D., et al.\ 2005, \aap, 440, 775 

\bibitem[Kneib et al.(2003)]{2003ApJ...598..804K} 
Kneib, J.-P., Hudelot, P., Ellis, R.~S., et al.\ 2003, \apj, 598, 804 

\bibitem[Lauer et al.(2014)]{2014ApJ...797...82L} 
Lauer, T.~R., Postman, M., Strauss, M.~A., Graves, G.~J., \& Chisari, N.~E.\ 2014, \apj, 797, 82 

\bibitem[Lin et al.(2009)]{2009ApJ...694..992L} 
Lin, Y.-T., Partridge, B., Pober, J.~C., et al.\ 2009, \apj, 694, 992 

\bibitem[Mann \& Ebeling(2012)]{2012MNRAS.420.2120M} 
Mann, A.~W., \& Ebeling, H.\ 2012, \mnras, 420, 2120 

\bibitem[Mantz et al.(2016)]{2016MNRAS.463.3582M} % WtG V
Mantz, A.~B., Allen, S.~W., Morris, R.~G., et al.\ 2016, \mnras, 463, 3582 

\bibitem[Martino et al.(2014)]{2014MNRAS.443.2342M} 
Martino, R., Mazzotta, P., Bourdin, H., et al.\ 2014, \mnras, 443, 2342 

\bibitem[Maughan et al.(2012)]{2012MNRAS.421.1583M} 
Maughan, B.~J., Giles, P.~A., Randall, S.~W., et al.\ 2012, \mnras, 421, 1583

\bibitem[McDonald et al.(2014)]{2014ApJ...794...67M} 
McDonald, M., Benson, B.~A., Vikhlinin, A., et al.\ 2014, \apj, 794, 67 

\bibitem[Miyazaki et al.(2018)]{2018PASJ...70S..27M} 
Miyazaki, S., Oguri, M., Hamana, T., et al.\ 2018, \pasj, 70, S27

\bibitem[Moretti et al.(2009)]{2009A&A...493..501M} 
Moretti, A., Pagani, C., Cusumano, G., et al.\ 2009, \aap, 493, 501 

\bibitem[Mushotzky et al.(2019)]{2019arXiv190304083M} 
Mushotzky, R.~F., Aird, J., Barger, A.~J., et al.\ 2019, Astro2020 Decadal Survey (arXiv:1903.04083).

\bibitem[Navarro et al.(1997)]{1997ApJ...490..493N} 
Navarro, J.~F., Frenk, C.~S., \& White, S.~D.~M.\ 1997, \apj, 490, 493 

\bibitem[Okabe \& Smith(2016)]{2016MNRAS.461.3794O} 
Okabe, N., \& Smith, G.~P.\ 2016, \mnras, 461, 3794 

\bibitem[Owen \& Ledlow(1997)]{1997ApJS..108...41O} 
Owen, F.~N., \& Ledlow, M.~J.\ 1997, \apjs, 108, 41 

\bibitem[Pacaud et al.(2007)]{2007MNRAS.382.1289P} 
Pacaud, F., Pierre, M., Adami, C., et al.\ 2007, \mnras, 382, 1289 

\bibitem[Pearson et al.(2017)]{2017MNRAS.469.3489P} 
Pearson, R.~J., Ponman, T.~J., Norberg, P., et al.\ 2017, \mnras, 469, 3489 

\bibitem[Pointecouteau et al.(2005)]{2005A&A...435....1P} 
Pointecouteau, E., Arnaud, M., \& Pratt, G.~W.\ 2005, \aap, 435, 1 

\bibitem[Planck Collaboration et al.(2011)]{2011A&A...536A...9P} %Lx-Y predic for REXCESS
Planck Collaboration, Aghanim, N., Arnaud, M., et al.\ 2011, \aap, 536, A9 

\bibitem[Planck Collaboration et al.(2012)]{2012A&A...543A.102P} 
Planck Collaboration, Aghanim, N., Arnaud, M., et al.\ 2012, \aap, 543, AA102 

\bibitem[Planck Collaboration et al.(2013)]{2013A&A...550A.129P} 
Planck Collaboration, Ade, P.~A.~R., Aghanim, N., et al.\ 2013, \aap, 550, A129 

\bibitem[Planck Collaboration et al.(2014)]{2014A&A...571A..29P} % il catalogo PZ1
Planck Collaboration, Ade, P.~A.~R., Aghanim, N., et al.\ 2014, \aap, 571, A29 

\bibitem[Planck Collaboration et al.(2016)]{2016A&A...594A..27P} % outlier
Planck Collaboration, Ade, P.~A.~R., Aghanim, N., et al.\ 2016, \aap, 594, A27 

\bibitem[Pratt \& Arnaud(2005)]{2005A&A...429..791P} 
Pratt, G.~W., \& Arnaud, M.\ 2005, \aap, 429, 791 

\bibitem[Pratt et al.(2009)]{2009A&A...498..361P} % scaling
Pratt, G.~W., Croston, J.~H., Arnaud, M., \& B{\"o}hringer, H.\ 2009, \aap, 498, 361 

\bibitem[Pratt et al.(2010)]{2010A&A...511A..85P} % entropy
Pratt, G.~W., Arnaud, M., Piffaretti, R., et al.\ 2010, \aap, 511, A85 

\bibitem[Rasia et al.(2004)]{2004MNRAS.351..237R} 
Rasia, E., Tormen, G., \& Moscardini, L.\ 2004, \mnras, 351, 237.

\bibitem[Renzini \& Andreon(2014)]{2014MNRAS.444.3581R} 
Renzini, A., \& Andreon, S.\ 2014, \mnras, 444, 3581 

\bibitem[Rines \& Diaferio(2006)]{2006AJ....132.1275R} 
Rines, K., \& Diaferio, A.\ 2006, \aj, 132, 1275 

\bibitem[Rossetti et al.(2017)]{2017MNRAS.468.1917R} 
Rossetti, M., Gastaldello, F., Eckert, D., et al.\ 2017, \mnras, 468, 1917 

\bibitem[Sanders et al.(2018)]{2018MNRAS.474.1065S} 
Sanders, J.~S., Fabian, A.~C., Russell, H.~R., \& Walker, S.~A.\ 2018, \mnras, 474, 1065 

\bibitem[Schechter(1976)]{1976ApJ...203..297S} 
Schechter, P.\ 1976, \apj, 203, 297

\bibitem[Serra et al.(2011)]{2011MNRAS.412..800S} 
Serra, A.~L., Diaferio, A., Murante, G., \& Borgani, S.\ 2011, \mnras, 412, 800 

\bibitem[Smith et al.(2001)]{2001ApJ...556L..91S} 
Smith, R.~K., Brickhouse, N.~S., Liedahl, D.~A., \& Raymond, J.~C.\ 2001, \apjl, 556, L91 

\bibitem[Stanek et al.(2006)]{2006ApJ...648..956S} 
Stanek, R., Evrard, A.~E., B{\"o}hringer, H., et al.\ 2006, \apj, 648, 956

\bibitem[Sun et al.(2009)]{2009ApJ...693.1142S} 
Sun, M., Voit, G.~M., Donahue, M., et al.\ 2009, \apj, 693, 1142 

\bibitem[Sun et al.(2011)]{2011ApJ...727L..49S} 
Sun, M., Sehgal, N., Voit, G.~M., et al.\ 2011, \apjl, 727, L49 

\bibitem[Takizawa et al.(2010)]{2010PASJ...62..951T} 
Takizawa, M., Nagino, R., \& Matsushita, K.\ 2010, \pasj, 62, 951

\bibitem[Vikhlinin et al.(2005)]{2005ApJ...628..655V} 
Vikhlinin, A., Markevitch, M., Murray, S.~S., et al.\ 2005, \apj, 628, 655 

\bibitem[Vikhlinin et al.(2006)]{2006ApJ...640..691V} 
Vikhlinin, A., Kravtsov, A., Forman, W., et al.\ 2006, \apj, 640, 691 

\bibitem[Vikhlinin et al.(2009)]{2009ApJ...692.1060V} 
Vikhlinin, A., Kravtsov, A.~V., Burenin, R.~A., et al.\ 2009, \apj, 692, 1060

\bibitem[Voit et al.(2005)]{2005MNRAS.364..909V} 
Voit, G.~M., Kay, S.~T., \& Bryan, G.~L.\ 2005, \mnras, 364, 909

\bibitem[Xu et al.(2018)]{2018A&A...619A.162X} 
Xu, W., Ramos-Ceja, M.~E., Pacaud, F., Reiprich, T.~H., \& Erben, T.\ 2018, \aap, 619, A162 

\bibitem[Walker et al.(2019)]{2019arXiv190304550W} 
Walker, S.~A., Nagai, D., Simionescu, A., et al.\ 2019, Astro2020 Decadal Survey (arXiv:1903.04550).

\bibitem[Willis et al.(2005)]{2005MNRAS.363..675W} 
Willis, J.~P., Pacaud, F., Valtchanov, I., et al.\ 2005, \mnras, 363, 675 


\end{thebibliography}
\end{document}